\documentclass{article}

\usepackage[utf8]{inputenc}

\usepackage{fix-cm,type1cm}

\usepackage{amsmath}

\usepackage{amsfonts}
\usepackage{xcolor}

\usepackage{graphicx}
\graphicspath{{./pictures/}}
\DeclareGraphicsExtensions{.eps,.eps.bz2}
\newcommand{\vc}[1]{\mbox{\boldmath\ensuremath{#1}}}  
\newcommand{\ts}[1]{\mbox{\boldmath\ensuremath{#1}}}  
\newcommand{\ud}{\,\mathrm{d}} 
\newcommand{\sgn}{\mathrm{sign}}
\newcommand{\ensavg}[1]{\langle #1 \rangle} 
\newcommand{\favg}[1]{\widehat{#1}} 
\newcommand{\fder}[1]{{#1}^\circ} 
\newcommand{\pavg}[1]{\overline{#1}} 
\newcommand{\pder}[1]{#1'} 
\newcommand{\vf}{\phi} 
\newcommand{\svf}{\phi_s} 
\newcommand{\fvf}{{\phi_f}} 
\newcommand{\svfcrit}{{\phi_{sc}}} 
\newcommand{\bdyMom}{{M_s^d}} 
\newcommand{\stress}{\ts \tau} 
\newcommand{\stresscmpt}{\tau} 
\newcommand{\sstress}{{\ts{\tau}_s}} 
\newcommand{\sstresscmpt}{{\tau_s}} 
\newcommand{\fstress}{{\ts{\tau}_f}} 
\newcommand{\fstresscmpt}{{\tau_f}} 
\newcommand{\fsrt}{{\ts{\dot \gamma}_f}} 
\newcommand{\ssrt}{{\ts{\dot \gamma}_s}} 
\newcommand{\vel}{\vc{u}} 
\newcommand{\svel}{{\vc{u}_s}} 
\newcommand{\fvel}{{\vc{u}_f}} 
\newcommand{\vvel}{{\vc{v}}} 
\newcommand{\wvel}{{\vc{w}}} 
\newcommand{\fvelu}{{u_f}} 
\newcommand{\svelu}{{u_s}} 
\newcommand{\prs}{p} 
\newcommand{\fprs}{{p_f}} 

\newcommand{\eps}{\varepsilon}
\newcommand{\be}{\begin{equation}}
\newcommand{\ee}{\end{equation}}
\newcommand{\bea}{\begin{eqnarray}}
\newcommand{\eea}{\end{eqnarray}}
\newcommand{\beas}{\begin{eqnarray*}}
\newcommand{\eeas}{\end{eqnarray*}}

\newcommand{\pt}{\partial_t}
\newcommand{\px}{\partial_x}

\newcommand{\py}{\partial_y}
\newcommand{\pyy}{\partial_{yy}}
\newcommand{\pz}{\partial_z}

\newcommand{\pxi}{\partial_{\xi}}
\newcommand{\pxixi}{\partial_{\xi\xi}}

\newcommand{\ph}{\phi_0\left(1/2\right)}
\newcommand{\pht}{\phi^2_0\left(1/2\right)}
\newcommand{\phc}{\phi^3_0\left(1/2\right)}
\newcommand{\Dar}{{\mathrm{Da}}}

\newcommand{\rf}[1]{\eqref{#1}}

\title{Models for the two-phase flow of concentrated suspensions}	 

\author{
  Tobias Ahnert\thanks{
  Institute of Mathematics, Technische Universit\"at Berlin, Strasse des 17. Juni 136, 10623 Berlin, Germany}, 
  Andreas M\"unch\thanks{Mathematical Institute, University of Oxford, Andrew Wiles Building, Oxford OX2 6GG, UK}, 
Barbara Wagner\thanks{
  Weierstrass Institute of Applied Analysis and Stochastics, Mohrenstrasse 39, 10623 Berlin, Germany}
}

\date{}

\begin{document}

\maketitle

\begin{abstract}
A new two-phase model for concentrated suspensions is derived that incorporates a constitutive law combining the rheology for non-Brownian suspension and  granular flow. The resulting model exhibits a yield-stress behavior for the solid phase depending on the collision pressure. This property is investigated for the simple geometry of plane Poiseuille  flow, where an unyielded or jammed zone of finite width arises in the center of the channel. For the steady states of this problem, the governing equations are reduced to a boundary value problem for a system of ordinary differential equations and the conditions for existence of solutions with jammed regions are investigated using phase-space methods. For the general time-dependent case a new drift-flux model is derived using matched asymptotic expansions that takes into account the boundary layers at the walls and the interface between the yielded and unyielded region.
The drift-flux model is used to numerically study the dynamic behavior of the suspension flow including the appearance and evolution of an unyielded or jammed region.
\end{abstract}


\section{Introduction}

Ever since the derivation of an effective viscosity for dilute suspensions by
Einstein \cite{Einstein1906} and its extensions by Batchelor \& Green
\cite{Batchelor1972}, there have been numerous investigations into the
rheological properties of suspensions. Since the experimental work by
{Gadala-Maria \& Acrivos}~\cite{Gadalamaria1980} and {Leighton \&
Acrivos}~\cite{Leighton1987}, the discovery of physical phenomena related to
shear-induced particle migration for concentrated suspensions from regions of
high to low stress has spurred theoretical investigations that led to
expressions for associated diffusive flux terms as well as drift-flux models,
see for example \cite{Leighton1987, Phillips1992, Chow1994, Morris1999}. 

Even though drift-flux models are quite popular and are frequently used as a transport mechanism \cite{Cook2008, Murisic2013, Zhou2005}, they sometimes predict unphysical migration behavior such as a sharp peak in the particle volume fraction profile in the center of flow through a channel, where the shear rate vanishes \cite{Nott1994}, whereas in experiments~\cite{Hampton1997} the concentration profile is in fact flattened there. While these issues were addressed by introducing a suspension temperature as a measure of particle velocity fluctuations \cite{Nott1994,Jenkins1990, Wylie2003}, or by slightly changing parameter values such as the exponent in the Krieger-Dougherty law or in the expressions for the relative suspension viscosity and the particulate phase pressure as the maximum packing fraction is approached \cite{Ramachandran2013}, it remains to understand how these models can be based on their underlying two-phase models. This has been particularly difficult for highly concentrated suspensions, when additional flow transitions, such as jammed states may occur. Some fundamental studies of these flow regimes have been presented in Cassar et al.~\cite{Cassar2005}, where it was shown that for highly dense suspensions of particles in a viscous liquid that is sheared at a rate $\dot \gamma$ under a confining pressure $p_p$ can be characterized by a single dimensionless control parameter, the ``viscous number'' $I_v=\eta_f\dot\gamma/p_p$, where $\eta_f$ is the fluid viscosity. These findings have been supported by experiments where the suspensions are  sheared with a constant particle pressure~\cite{Boyer2011}. Their results show that, indeed, the friction and volume-fraction law collapse onto universal curves when expressed in terms of the dimensionless number $I_v$. By including hydrodynamic contributions, Boyer et al.\ propose a model for the whole range of $I_v$, which has been discussed by \cite{Bruyn2011}, and also by Trulsson et al. \cite{Trulsson2012}. An earlier review of stress terms for dense suspensions can be found in \cite{Stickel2005} and more recently for the special case of Houska fluids analytical solutions for unidirectional pipe flow have been derived in \cite{Ahmadpour201323}, while other approaches such as by Quemada in 
\cite{Quemada1, Quemada2, Quemada3, Quemada4} introduce structural models for concentrated suspensions, where shear-dependent effective volume fractions are introduced to take account of structures of the flow, such as clusters into account.

Boyer et al. formulate their model in a form, where the shear stress and particle pressure are expressed in terms of the strain rate and the volume fraction. Their expressions for the shear and normal viscosities are similar to the ones found in Morris \& Boulay~\cite{Morris1999}, and  also Miller et al.~\cite{Miller2009}, who investigated more general curvilinear flows, where the migration behavior was accommodated by allowing for anisotropy in the normal stresses.  

In section~2, we derive a new two-phase model for non-homogeneous shear flows that captures the flow properties of non-Brownian dense suspensions by including the  constitutive equations proposed by Boyer et al. \cite{Boyer2011}. The derivation is based on the averaging framework as given in Drew \& Passman \cite{Drew1983,Drew1999} and is formulated for a general three dimensional flow.
For the remainder of the article, 
we focus on the pressure-driven plane Poiseuille flow as our model example for non-constant shear flows and investigate the flow behavior predicted by the two-phase model as the particle volume fraction is varied. 

In section 3, we first consider stationary solutions for the plane Poiseuille flow,
for which the model reduces to a bound\-ary-value problem for a system of ordinary differential equations. Using  phase-space methods we reveal the existence of solutions that show an unyielded
region similar due to a yield-stress condition for the solid phase that is similar to
the condition in Bingham-type flows. In this region, located at the center of
the channel, the solid volume fraction is at
its maximum and the solid phase has jammed to form a porous matrix through which the fluid
phase can still flow through. We study the dependence of the width of the unyielded i.e.\ jammed region (for the solid phase) and the flow fields for both phases upon varying the flow parameters. 
We then show that for typical ranges of the parameter $\Dar=L/K_p\ll O(1)$ the flow field develops boundary layers at the channel walls and at the interface between the unyielded and yielded region, where $L$ is the characteristic scale of the channel width and $K_p$ is proportional to the particle size. Using matched asymptotic analysis as $\Dar\to\infty$ we obtain an expression for the flow field. 

In section 4, we generalize this analysis and for the first time present a systematic asymptotic derivation of a time-dependent drift-flux model via matched asymptotic expansions. Our numerical simulations of the drift-flux model captures the emergence of a jammed region and the evolution of the flow and phase field into a stationary state.

In section 5, we summarize our results and give an outlook on future directions for research.

\section{Formulation of the two-phase model}
\label{sec:multiphase_model}

We consider two inert phases, a solid phase of particles suspended in a liquid phase where we denote with $k \in \{s,f\}$ the solid phase by $s$ and the  liquid phase by $f$. Inside each phase the balance equations for mass and momentum
\begin{subequations} 
\begin{align}\label{eqn:balance}
 \partial_t \rho_k + \nabla \cdot (\rho_k \vel_k) &= 0 \\
 \partial_t (\rho_k \vel_k) + \nabla \cdot (\rho_k \vel_k \otimes \vel_k) - \nabla \cdot \ts{T}_k - \vc{f}_k &= \vc{0}
\end{align}
are satisfied together with the two jump conditions (see e.g. \cite{Ishii2011}) \begin{align}\label{eqn:jump_cond_def1}
\sum_{k} \rho_k (\vel_k - \vel_i) \cdot \vc{n}_k &= 0 \\
\sum_{k} \rho_k \vel_k(\vel_k - \vel_i) \cdot \vc{n}_k - \ts{T}_k \cdot \vc{n}_k &= \vc{0}, \label{eqn:jump_cond_def2}
\end{align}
\end{subequations}
at the interfaces of the phases with $\vc{n}_k$ denoting the unit normal pointing out of phase $k$ and $\vc{u}_i$ is the interface velocity.
The quantities $\rho$, $\vel$, $\bf T$ and $\bf f$ denote density, velocity, stress tensor and body force density in each phase, respectively. At an interface $\vel_k$ is defined as
$$\vel_k(\vc{x}^*,t) \equiv \lim_{\vc{x} \to \vc{x}^*; \vc{x} \in K} \vel(x,t),$$ 
where $K$ denotes the set of points occupied by phase $k$, and similarly for the other quantities.

In deriving a two-phase model, essentially three different averaging approaches have been pursued in the literature. The volume average, the time average and the ensemble average (sometimes also called statistical average). Although all three produce similar balance equation for the phases their derivation and closure is distinct. Besides the ensemble averaging developed by Drew \& Passman in \cite{Drew1999} and \cite{Drew1983}, volume averaging is treated in  Kolev \cite{Kolev2005} and Whitaker \cite{Whitaker1998} and time averaging in Ishii et al. \cite{Ishii2011}. 

For the derivation of our two-phase model we follow the mathematical framework by Drew \cite{Drew1983} and Drew and Passman \cite{Drew1999} and introduce a component indicator function 
\begin{equation}
 X_k(\vc{x},t) =
\begin{cases}
 1, &\text{ if } (\vc{x},t) \in K \\
 0, &\text{ if } (\vc{x},t) \not\in K
\end{cases}
\end{equation}
with $K$ the set of states of the $k$-th-phase to define the averaged quantities
\be
\vf_k \equiv \ensavg{X_k}, 
\quad 
\pavg{\rho}_k \equiv \frac{\ensavg{X_k \rho}}{\vf_k},  
\quad 
\favg{\vel}_k \equiv \frac{\ensavg{X_k \rho \vel}}{\vf_k \pavg{\rho}_k}, 
\quad 
\pavg{p}_k \equiv \frac{\ensavg{X_k p}}{\vf_k},  
\quad 
\pavg{\ts{\tau}}_k \equiv -\frac{\ensavg{X_k \ts{\tau}}}{\vf_k}
\ee
which denote the volume fractions, the averaged densities, velocities,  pressures and deviatoric stresses, respectively. For these quantities we then derive   the following balance equations 
\begin{subequations} 
\label{eqn:govern_dimens_system}
\begin{align}
  \partial_t \svf + \nabla \cdot (\svf \favg{\vel}_s) &= 0, \label{conti_fluid_ssm} \\
  \partial_t \fvf + \nabla \cdot (\fvf \favg{\vel}_f) &= 0, \\
\pavg{\rho}_s \partial_t (\svf \favg{\vel}_s) + \nabla \cdot (\svf \pavg{\rho}_s \favg{\vel}_s \otimes \favg{\vel}_s) \quad \\
- \nabla \cdot (\svf \pavg{\stress}_s) + \nabla (\svf \pavg{\prs}_s) &= \bdyMom + \pavg{\prs}_f \nabla \svf, \\
\pavg{\rho}_f \partial_t (\fvf \favg{\vel}_f) + \nabla \cdot (\fvf \pavg{\rho}_f \favg{\vel}_f \otimes \favg{\vel}_f) \quad \label{ssm_momeqn_fluid} \\
- \nabla \cdot (\fvf \pavg{\stress}_f) + \nabla (\fvf \pavg{\prs}_f) &= -\bdyMom + \pavg{\prs}_f \nabla \fvf.
\end{align}
\end{subequations}
The detailed derivation is given in appendix \ref{app:derivation}, where we note that in the present study we also have neglected the Reynolds stresses (see Drew \cite{Drew2001}) and their possible impact on dispersion and boundary layers. 


\subsection{Constitutive equations for a dense suspension}
To close the model for the flow in the bulk, we need to specify constitutive equations besides the assumptions already made. Essentially we need four relations for the pressure difference and stress between the phases $\pavg{\prs}_s-\pavg{\prs}_f$ and $\bdyMom$, and for the stresses in each phase, $\pavg{\stress}_f$ and $\pavg{\stress}_s$. In addition, to simplify notation, we set $\phi \equiv \phi_s$.
\renewcommand{\svf}{{\phi}}
\renewcommand{\fvf}{{(1 - \phi)}}

For the momentum transfer 
\begin{equation}\label{kcdrag}
 \bdyMom = \frac{\mu_f \svf^2}{K_p \fvf}(\favg{\vel}_f - \favg{\vel}_s),
\end{equation} 
we have used the Kozeny-Carman equation, with $K_p$ depending on the particle
diameter, $K_p\propto a^2$, see e.g.\ 
\cite{Brennen2005}.
We note that more general closures could have been used for a wider range of $\phi_s$, in particular
for the medium range, see for example \cite{Garrido200357} and references therein and also \cite{Ahnert2015}.

The constitutive law for the remaining quantities extend the model for dense suspensions given by Boyer et al.\ \cite{Boyer2011} for shear flow to a general flow situation. We state it in terms of the (weighted) solid contact pressure, defined here as
\begin{equation}\label{pcdef}
  p_c\equiv \svf (\pavg{\prs}_s - \pavg{\prs}_f),
\end{equation}
and the shear rate tensors for each phase
\begin{align}
  \fsrt & \equiv [\nabla \favg{\vel}_f + (\nabla \favg{\vel}_f)^T], \qquad
  \ssrt  \equiv [\nabla \favg{\vel}_s + (\nabla \favg{\vel}_s)^T].
\end{align}

For the liquid phase stress, we have
\begin{subequations}\label{rheo}
\begin{equation}
\pavg{\stress}_f = \mu_f \fsrt + (\mu^* - \frac{2}{3}\mu_f)(\nabla \cdot \favg{\vel}_f) \ts{I},
\label{taufrel}
\end{equation}
where $\mu_f$ denotes the viscosity of the pure liquid. The second term in the relation for $\pavg{\stress}_f$ will be dropped by setting the bulk viscosity $\mu^* = \frac{2}{3} \mu_f$. This particular choice has no influence on the solutions, since generally the bulk viscosity vanishes for stationary flows and does not appear in the leading order approximation of the derived drift-flux term.

For the solid phase, we need to consider two cases for the constitutive law: 
\\
Either $|\ssrt|>0$, then
\begin{align}
 \pavg{\stress}_s &= \mu_f \eta_s(\svf) \ssrt, \label{rheotaus}\\
 p_c &= \mu_f \eta_n(\svf) |\ssrt|, \label{eqn:ssm_p_c_def}
 \end{align}
with
\begin{align}
   \eta_s(\svf) &= 1+\frac{5}{2} \frac{\svfcrit}{\svfcrit - \svf}
+ \mu_c(\svf) \frac{\svf}{(\svfcrit - \svf)^2}, \qquad\label{etas_nondim} \\
  \mu_c(\svf)  &= \mu_1 + \frac{\mu_2 - \mu_1}{1 + I_0 \svf^2 (\svfcrit - \svf)^{-2}}, \label{muc_nondim}\\
  \eta_n(\svf) &= \left(\frac{\svf}{\svfcrit - \svf}\right)^2, \label{etan_nondim}
 \end{align}
where for tensors $\ts{a}$ we define the norm as $|\ts{a}| = (\frac{1}{2} \sum_{j,k} |a_{jk}|^2)^{1/2}$. The parameters $\mu_2\geq\mu_1>0$, $I_0\geq 0$ characterize the contact contribution in the expression for $\eta_s$, and $\svfcrit$ is the maximum volume fraction for the solid phase,
which is attained when the solid phase jams.
We note from the experimentally fitted laws in Boyer et al. \cite{Boyer2011}, \eqref{etas_nondim}-\eqref{etan_nondim} is found from the friction law for dense suspensions $\mu_s= \mu_1 +(\mu_2-\mu_1)/(1+I_0/I_{\nu}) +I_{\nu}+5/2\phi_{sc}I_{\nu}^{1/2}$ scaled by $I_{\nu}=[(\phi_{sc}-\svf)/\svf]^2$.   

For the other case $\ssrt=0$ we require
\begin{equation}
\svf=\svfcrit,
\end{equation} 
while $\pavg{\stress}_s$, $\pavg{\prs}_s$ and $\pavg{\prs}_f$ are left unspecified, except for the constraint
\begin{equation}\label{constraint}
|\svf \pavg{\stress}_s|
\leq \mu_1 p_c.
\end{equation}
\end{subequations}

Conversely, if \eqref{constraint} is satisfied, then it follows from
equations \eqref{rheotaus}-\eqref{etan_nondim} that $|\ssrt|$ cannot be positive.
Thus, if the collision pressure 
$p_c$ is finite, our model for concentrated suspensions is capable of
exhibiting regions where the solid phase is jammed whenever $|\pavg{\stress}_s|$ 
drops below a certain yield stress. In the jammed region the solid phase flow
is plug-like. This is similar to the plug-like flow in a Bingham model, which, 
however, describes single phase rheology with a constant yield stress.

The jammed regions are separated from the regions where $|\ssrt|>0$ by
yield surfaces.
Across a yield surface,  we require $\svf$, $\favg{\vel}_f$, $\favg{\vel}_s$, $(-\pavg{\prs}_f\ts{I}+\pavg{\stress}_f) \cdot\vc{n}_{y}$, $(-\pavg{\prs}_s\ts{I}+\pavg{\stress}_s) \cdot\vc{n}_{y}$ and $|\ssrt|$ to be continuous, where 
$\vc{n}_{y}$ denotes the unit normal vector to the surface.

\subsection{Non-dimensionalization}

We introduce characteristic scales via
\begin{align}
x &= L x' ,&
y &= L y' ,&
z &= L z' ,& 
t &= \frac{L}{U} t', & \\
\vel_k &= U \vel_k',&
p_k &= \frac{U \mu_f}{L} p_k', &
\stress_k &= \frac{U \mu_f}{L} \stress_k', &
\end{align}
for $k=s$, $f$.
After non-dimensionalization, we drop the primes and also the bars and hats indicating averaging, and obtain the system
\begin{subequations}\label{bmod:nondimfinal}
\begin{align}
 \pt \fvf + \nabla \cdot (\fvf \fvel) &= 0, \label{bmod:nondimfinala}\\
 \pt \svf + \nabla \cdot (\svf \svel) &= 0, \label{bmod:nondimfinalb}\\
{\mathrm{Re}} [\pt (\fvf \fvel) + \nabla \cdot (\fvf \fvel \otimes \fvel)] \quad
\label{bmod:nondimfinalc}\\ \notag
- \nabla \cdot (\vf_f \fstress) + \fvf \nabla \fprs &= -\Dar\, \frac{\svf^2}{\fvf}(\fvel - \svel),
\\
\frac{\mathrm{Re}}{r} [\pt (\svf \svel) + \nabla \cdot (\svf \svel \otimes \svel)] \quad
\label{bmod:nondimfinald}\\ \notag 
- \nabla \cdot (\svf \sstress) + \svf \nabla \fprs + \nabla p_c 
&= \Dar\,  \frac{\svf^2}{\fvf}(\fvel - \svel).
\end{align}
\end{subequations}
Three dimensionless numbers appear here, namely the Reynolds number $\mathrm{Re} = {U L \rho_f}/{\mu_f}$, the Darcy number $\Dar = {L^2}/{K_p}$
and the density ratio $r = {\rho_f}/{\rho_s}$. We focus on the case where liquid and solid phases are density matched and set $r=1$. 

The non-dimensional versions of the constitutive equations for the rheology 
are now as follows: For the liquid phase, we have
\begin{subequations}\label{rheond}
\begin{align}
 {\fstress} &= \fsrt. 
\label{taufrel_nondim}\\
\intertext{For the solid phase, either $|\ssrt|>0$, then}
 {\sstress} &=  \eta_s(\svf) \ssrt, \label{tausrel_nondim} \\
\label{rheondc}
 p_c &=  \eta_n(\svf) |\ssrt|,
 \end{align}
\end{subequations} 
with \eqref{etas_nondim}-\eqref{etan_nondim}; or $\ssrt=0$, and then we require
$$\svf=\svfcrit$$
and
$$|\svf \sstress|\leq \mu_1 p_c.$$
The continuity conditions across yield surfaces
carry over from the dimensional equations and also
the parameters, $\mu_1$, $\mu_2$ and $I_0$ and $\svfcrit$, which
were non-dimensional to begin with.

\section{Plane Poiseuille flow}\label{sec:channel_flow}

\begin{figure}
\centerline{\includegraphics[width=0.8\linewidth]{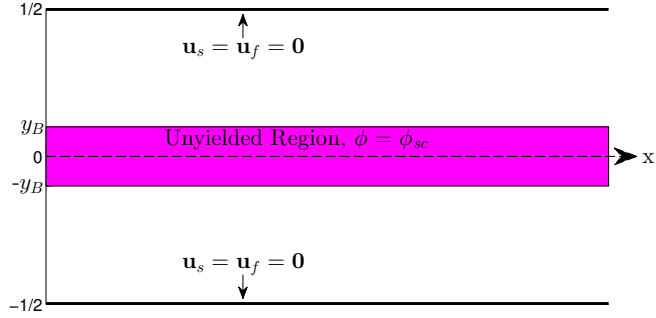} }
 \caption{Sketch of the flow situation in a channel.}\label{fig:sc}
\end{figure}

It is instructive to investigate the properties of the model \eqref{bmod:nondimfinal} for one of the classical flow situations, namely, plane Poiseuille or channel flow, which we think is the simplest flow geometry to exhibit the emergence of a jammed or unyielded region. 
The dimensions of the channel are $0<x<x_e$ and $-\frac12<y<\frac12$, where we have
used the channel with for the length scale $L$, and prescribe for the inlet conditions at $x=0$
\begin{equation}\label{inlet}
\svf=\phi_{in},\quad
\fvel=\begin{pmatrix}
u_{f,in}\left(
\frac14-y^2
\right) 
\\
0
\end{pmatrix}, \quad
\svel=\begin{pmatrix}
u_{s,in}\left(\frac14-y^2\right) \\
0
\end{pmatrix}
\end{equation}
and for the outlet condition at $x=x_e$
\begin{equation}
{\bf n}\cdot(p_s \ts{I} + \svf\eta_s(\nabla \svel)^T) = 0,\quad {\bf n} \cdot(p_f \ts{I} + \fvf \eta_s(\nabla \svel)^T)=0.
\end{equation}
Moreover, we can set $u_{f,in}=1$ by appropriately choosing the velocity scale $U$. 
In addition, we only consider rectilinear flow, so the inertial terms vanish and we obtain for the bulk equations
\begin{subequations}
\begin{align}
\pt\fvf + \nabla \cdot (\fvf \fvel) &= 0, \\
\pt\svf + \nabla \cdot (\svf \svel) &= 0, \\
-\nabla\cdot(\fvf \fstress) + \fvf \nabla \fprs &= -\Dar\frac{\svf^2}{\fvf} (\fvel-\svel), \\
-\nabla\cdot(\svf \sstress) + \svf \nabla \fprs + \nabla p_c &= \Dar\frac{\svf^2}{\fvf} (\fvel-\svel),
\end{align}
\end{subequations}
where 
\begin{subequations}
\begin{align}
\fstress &= \fsrt \label{eqn:fstress_def} \\
\left|\svf \sstress\right| &\leq \mu_1 p_c,& \svf &= \svfcrit & 
\mbox{if } |\ssrt| &= 0 \label{eqn:sstress_def_solid} \\
\sstress &=\eta_s(\svf)\ssrt,& p_c &= \eta_n(\svf)|\ssrt|  &
\mbox{if } |\ssrt| &\neq 0. \label{eqn:sstress_def_fluid}
\end{align}
\end{subequations}
At the channel walls we assume the no-slip conditions
\begin{equation}
\svel=\vc{0},\qquad \fvel = \vc{0}. 
\end{equation}
For the two-phase model at hand, it turns out to be advantageous to formulate the problem in terms of the flow variables 
\begin{equation}
\vvel \equiv \fvf \fvel + \svf \svel, \qquad \wvel \equiv \fvel - \svel .
\end{equation}
In these variables, noting that 
$\vvel + \svf \wvel = \fvel $, $\vvel-\fvf \wvel = \svel $ the problem can be written as 
\begin{subequations}
\begin{align}
\nabla \cdot \vvel &= 0 \label{channela}\\
\pt \svf + \nabla \cdot (\svf \vvel - \svf(1-\svf) \wvel) &= 0 \label{channelb}\\
-\nabla\cdot \left((1-\svf) \fsrt \right) + (1-\svf)\nabla \fprs &= -\Dar \frac{\svf^2}{1-\svf} \wvel \label{channelc}\\
%
-\nabla\cdot(\svf \sstress) + \svf \nabla \fprs + \nabla p_c &= \Dar\frac{\svf^2}{\fvf} \wvel
\label{channeld}
\end{align}
where $\sstress$ satisfies the constitutive law \eqref{eqn:sstress_def_solid},
\eqref{eqn:sstress_def_fluid}.
At the walls $y=\pm \frac{1}{2}$, the no-slip conditions are
\begin{equation}
 \vvel=\vc{0}, \qquad \wvel=\vc{0}. \label{channele}
\end{equation}
\end{subequations}

\subsection{Formulation of the stationary problem}

For the system \eqref{channela}-\eqref{channele} we now derive conditions for the existence of stationary two-dimensional solutions where all quantities, except for the pressure, depend only on $y$,  
\begin{align}
	\phi &= \phi(y),& \vvel &= \vvel(y),& \wvel &= \wvel(y),\\
	\fstress &= \fstress(y),& \sstress &= \sstress(y),& \fprs &= \fprs(x,y).
\end{align}

The combination of no-slip boundary conditions \eqref{channele} with \eqref{channela},  \eqref{channelb}
yields (if $v_1$, $v_2$ and $w_1$, $w_2$ denote the components of the vectors
$\vvel$ and $\wvel$, respectively)
\begin{align}
 v_2 &= 0, & w_2 &= 0,
\end{align}
therefore
\begin{subequations}
\begin{align}
 \ssrt &= 
  \begin{pmatrix}
  0 & \partial_y (v_1 - (1-\phi) w_1) \\
  \partial_y (v_1 - (1-\phi) w_1) & 0
  \end{pmatrix},
\label{eqn:ssrt}\\
\fsrt &= 
\begin{pmatrix}
 0 & \partial_y (v_1 + \phi w_1) \\
 \partial_y (v_1 + \phi w_1) & 0
\end{pmatrix}.
\end{align}
\end{subequations}
The momentum balances \eqref{channelc} and \eqref{channeld} become, 
in components,
\begin{subequations}
\begin{align}
-\partial_y\left((1-\phi)\py (v_1+\phi w_1\right)+(1-\phi)\partial_x p_f
&=-\Dar\frac{\phi^2}{(1-\phi)} w_1,\label{eqn:mbfx}\\
\partial_y p_f &=0,\label{eqn:mbfy}\\
-\partial_y\left(\phi \tau_{s12}\right)+\phi\partial_x p_f+\partial_x p_c 
&=\Dar\frac{\phi^2}{(1-\phi)} w_1,\label{eqn:mbsx}\\
-\py \left(\phi \tau_{s22}\right)+\phi \partial_y p_f +\partial_y p_c&=0.
\label{eqn:mbsy}
\end{align} 
\end{subequations}
From \eqref{eqn:mbfy} we conclude that $p_f=p_f(x)$ is a function of $x$
only, and the same is true for $\phi \tau_{s22}+\partial_y p_c=c_1(x)$. The
momentum balance requires continuity of stresses and hence these two relations
must be satisfied even across yield surfaces. If we assume that the solid is
not stagnant everywhere, that is, \eqref{eqn:sstress_def_fluid} holds for
some $y$, we can deduce (because of \eqref{eqn:ssrt}) that $\tau_{s22}=0$
there and hence $p_c=c_1(x)$ and $\tau_{s22}=0$ everywhere. Using now the
second condition in \eqref{eqn:sstress_def_fluid} we see that, because the
right hand side depends only on $y$, $p_c$ must in fact be a constant,
which is free and thus acts as an additional parameter.
Using this in \eqref{channelc}, \eqref{channeld} and adding the two equations gives
\begin{equation}
\px \fprs = \py \stresscmpt_{12},
\end{equation}
where we have introduced the total stress
$\stress\equiv(1-\phi)\fstress+\phi\,\sstress
=(1-\phi)\fsrt+\phi\,\sstress$.
%
%
%
Since the left hand side only
depends on $x$ and the right hand side only on $y$, 
both have to be constant, and we get
\begin{subequations}
\begin{align}
 \fprs(x) &= p_1 x+p_0, 
\intertext{where $p_0$ is a constant of integration, 
which by a choice of origin, we can assume, without loss of 
generality, to be zero, and}
\stresscmpt_{12}(y) &= p_1 y.
\label{eqn:tube_ana_ssm_total_stress}
\end{align}
\end{subequations}
Here and from now on, we will only look at the case of solutions
with velocities and volume fractions that are symmetric
with respect to $y=0$, so that we have in particular set
the integration constant that would normally appear in
\eqref{eqn:tube_ana_ssm_total_stress} to 0 and will only consider one
half of the channel, $0\leq y\leq 1/2$.
Moreover, we assume that we have at most one unyielded region $0\leq y\leq y_B$
that is located
at the center of the channel and ends at $y_B$, $0\leq y_B\leq1/2$, which is an
unknown of the problem.   


Overall we get the system: For $y \in [y_B; 1/2]$, $\phi$, $\sstresscmpt_{12}$, $\fstresscmpt_{12}$, $v_1$ and $w_1$ satisfy
\begin{subequations}
\label{eqn:reduced_channel_ssm_with_yb}
\begin{align}
\py((1-\phi) \fstresscmpt_{12}) &= (1-\phi) p_1 +\Dar\,\frac{\phi^2}{1-\phi} w_1,
\label{eqn:reduced_channel_ssm_with_yb:a}\\
\svf \fstresscmpt_{12} &= \py(v_1+\phi w_1), 
\label{eqn:reduced_channel_ssm_with_yb:ab}\\
\svf \sstresscmpt_{12} &= p_1 y - (1-\phi) \fstresscmpt_{12}, \label{eqn:reduced_channel_ssm_with_yb:b}\\
\py w_1 &= \fstresscmpt_{12} - \frac{\sstresscmpt_{12}}{\eta_s(\phi)} \label{eqn:reduced_channel_ssm_with_yb:c}, \\
p_c &= \eta_n(\phi) |\partial_y (v_1 - (1-\phi) w_1)| . \label{eqn:reduced_channel_ssm_with_yb:d}
\end{align}
In the unyielded region $y\in[0;y_B[$, equations
\eqref{eqn:reduced_channel_ssm_with_yb:a}-\eqref{eqn:reduced_channel_ssm_with_yb:b}
stay the same, but the two remaining ones are replaced by 
\begin{align}
 \partial_y (v_1 - (1-\phi) w_1) = 0 
\qquad \text{ and } \qquad
\phi &= \svfcrit,
\label{eqn:tube_ana_ssm_svel_solid_def}
\end{align}

The boundary conditions are the no-slip
\begin{align}
v_1 & = 0, \quad w_1 = 0,  \qquad  \text{ at } y=1/2,\label{liqfixedbc}
\end{align}
and symmetry conditions
\begin{align}
\partial_y v_1 &=0, \quad \partial_y w_1 = 0,\qquad  \text{ at } y=0. \label{symbc} 
\end{align}
In case the unyielded region fills up the whole channel, i.e. $y_B = 1/2$, the no-slip boundary conditions together with \eqref{eqn:tube_ana_ssm_svel_solid_def} gives $v_1 - (1-\phi) w_1) = 0$. Then  \eqref{eqn:reduced_channel_ssm_with_yb:a} becomes the Brinkman equation, c.f. \cite{Brinkman1949}.
For the yield surface at $y = y_B$ we demand the continuity conditions
\begin{align}
[ \sstresscmpt_{12}]_-^+ &= 0,& 
[ \fstresscmpt_{12}]_-^+ &= 0,&
[v_1]_-^+ &= 0, \\
[w_1]_-^+ &= 0,&
[\phi]_-^+ &= 0, \label{eqn:reduced_channel_ssm_with_yb:l}
\end{align}
\end{subequations}
where we denote $[g]_-^+=\lim_{y\searrow y_B} g - \lim_{y\nearrow y_B} g$.
We remark that these conditions are not all independent, as, for example,
the second condition in \eqref{symbc} can be obtained from
the first via \eqref{eqn:tube_ana_ssm_svel_solid_def}, and the continuity of one of the stresses in \eqref{eqn:reduced_channel_ssm_with_yb:l}
implies the other via \eqref{eqn:reduced_channel_ssm_with_yb:b}.

Notice that \eqref{eqn:sstress_def_fluid} applies in the region
$[y_B; 1/2]$ where $\ssrt>0$, so that if $y_B<1/2$ (i.e.\ excluding the special
case where the entire channel is jammed with $\phi=\phi_{sc}$), then $p_c=0$
implies $\phi=0$. 

Notice that if $p_c=0$, then \eqref{eqn:sstress_def_fluid} implies that
$\phi=0$ in the region $[y_B; 1/2]$.  This would mean that all particles have
moved to the unyielded region and, unless we are in the special case where
$y_B=1/2$ and hence $\phi=\phi_{sc}$ everywhere, have left a clear liquid phase
behind. This is unplausible and certainly not what is observed in experiments,
e.g.\ in \cite{Hampton1997}, and it is not the type of solution that arises
from a homogeneous initial state in the time-dependent version of the equations
discussed in section~\ref{subsec:ad}, see
fig.~\ref{fig:driftfluxouter_asderived}.  We therefore assume $p_c>0$.
Then, we can remove $p_c$ from the equations by rescaling
\begin{equation}\label{pcrescaling1}
\sstresscmpt_{12} =p_c \tilde{\stresscmpt}_{s12}, \quad 
\fstresscmpt_{12} =p_c \tilde{\stresscmpt}_{s12},\quad
p_1 = p_c \tilde{p}_1,\quad
v_1 = p_c \tilde{v}_1,\quad
w_1 = p_c \tilde{w}_1.
\end{equation}
The fact that $p_c$ can be scaled out of the problem
in this way implies that the width
of the unyielded region i.e.\ $y_B$ does not depend on $p_c$, as was
reported in \cite{Isa2007}. 
We note that in conventional Herschel-Bulkley models, which are also able to capture yield
stress and shear-thinning, the unyielded region would change with $p_c$.

\subsection{Phase space analysis}

We now derive conditions for the existence of solutions to system  \eqref{eqn:reduced_channel_ssm_with_yb}. For this, it is convenient to reduce the system into a second order, non-autonomous system of ordinary differential equations for $w \equiv w_1$ and $\phi$. 
For convenience, we also introduce the notation
$\fvelu \equiv u_{f1}=v_1 + (1-\phi) w_1$ and 
$\svelu \equiv u_{s1}=v_1 - (1-\phi) w_1$ for the first components of
$\fvel$ and $\svel$, respectively.

We first note that in the fluid region $y \in [y_B; 1/2]$ combining the definition of the solid stress \eqref{eqn:sstress_def_fluid} and \eqref{eqn:reduced_channel_ssm_with_yb:d} yields
\begin{align}
	\phi \sstresscmpt_{12} = \vf \eta_s \py \svelu = \frac{\vf \eta_s }{\eta_n} \sgn(\py \svelu) = -\frac{\vf \eta_s }{\eta_n} \sgn(y),
	\label{eqn:taus_to_N}
\end{align}
Here we have used that $\sgn(\py \svelu) = \sgn(\sstresscmpt_{12}) = -\sgn(y)$, where we recall that due to \eqref{eqn:tube_ana_ssm_total_stress} the total stress is just a linear function of $y$. 
\begin{subequations}
\label{ODEformulation}
Then using \eqref{eqn:reduced_channel_ssm_with_yb:b} in \eqref{eqn:reduced_channel_ssm_with_yb:a} and \eqref{eqn:taus_to_N} yields
\begin{align}
	\py N(\vf) &= -\vf p_1 + \Dar \frac{\vf^2}{1 - \vf} w,
	\label{eqn:Ny_phi_w}
\end{align}
which will be used as an equation for the solid volume fraction. We get an equation for $w$ by combining \eqref{eqn:reduced_channel_ssm_with_yb:b} and \eqref{eqn:reduced_channel_ssm_with_yb:c} to give
\begin{align}
	\py w &= \frac{p_1 y + N(\vf)}{1 - \vf} + \frac{1 }{\eta_n(\vf)}.
	\label{eqn:wy_phi}
\end{align}
The function $N$ is given by
\begin{equation*}
N(\phi)\equiv\frac{\phi\eta_s(\phi) }{\eta_n(\phi)}.
\end{equation*}

In the unyielded region $y \in [0; y_B[$ we already know 
\begin{align}
	\phi = \svfcrit
\end{align}
and since $\py \svelu = 0$, we have $ \fstresscmpt_{12} = \py \fvelu = \py w$, which together with \eqref{eqn:reduced_channel_ssm_with_yb:a} is
\begin{align}
	\pyy w &= p_1 + \Dar \frac{\svfcrit^2}{(1-\svfcrit)^2} w.
	\label{wplug}
\end{align}

At the channel wall and center, we have the boundary conditions
\begin{align}
w&=0 & \text{ at } y &= 1/2, \label{eqn:w_no_slip_bc} \\
\partial_y w &= 0 & \text{ at } y &= 0, \label{wplugbc}
\end{align}
and at the yield surface,
\begin{equation}\label{ysc}
\phi=\svfcrit, \quad
[w]_-^+=0, \quad
[w_{y}]_-^+=0, \quad
\text{at } y= y_B.
\end{equation}
\end{subequations}
The problem for $w$ in the unyielded region, \eqref{wplug} and \eqref{wplugbc},
can now be solved explicitly. For $\Dar > 0$, we have
\begin{equation}
w=\alpha_1 \cosh\left(\frac{\Dar^{1/2}\svfcrit}{1-\svfcrit}y\right)-
\frac{(1-\svfcrit)^2}{\Dar\,\svfcrit^2 }p_1,
\end{equation}
where $\alpha_1$ is a constant of integration. We can use this 
in the last two conditions in \eqref{ysc} to get
\begin{align}
\partial_y w = \left(w+  \frac{(1-\svfcrit)^2}{\Dar\,\svfcrit^2 } p_1\right)
\frac{\Dar^{1/2}\svfcrit}{1-\svfcrit} 
\tanh\left(\frac{\Dar^{1/2}\svfcrit}{1-\svfcrit}y_B\right),  \text{ at } y = y_B
\end{align}
and from this a new formulation of the free boundary condition
\begin{subequations}
\begin{align}
\phi &= \svfcrit, \label{freebc_a}\\
w = W(y_B) &\equiv 
\frac{p_1 y_B +  \mu_1} 
{ \Dar^{1/2}\svfcrit 
\tanh\left(\frac{\Dar^{1/2}\svfcrit}{1-\svfcrit}y_B\right)}  -\frac{(1-\svfcrit)^2}{\Dar\,\svfcrit^2 } p_1, 
\quad \text{at } y= y_B \label{freebc}.
\end{align}
\end{subequations}
We have thus reduced the problem to a free boundary value problem for a second order system of ordinary differential equations \eqref{eqn:Ny_phi_w}, \eqref{eqn:wy_phi} with a condition \eqref{eqn:w_no_slip_bc} at the fixed boundary and two at the free boundary \eqref{freebc_a}, \eqref{freebc}. 

 
This free boundary value problem
contains the parameters $\Dar$, $\mathrm{\svfcrit}$, $\mathrm{p_1}$, $\mathrm{\mu_1}$, $\mathrm{\mu_2}$, $\mathrm{I_0}$. The critical volume fraction $\svfcrit$ is typically chosen between $0.63$ - $0.68$ (volume fraction at maximum random packing). The channel pressure gradient value $p_1$ will always be negative and for concentrated suspensions $\Dar$, which proportional $(L/a)^2$ is typically quite large, see e.g.  \cite{Nott1994,Morris1999}. The three parameter $\mu_1, \mu_2$ and $I_0$ are material parameters. In our study we fix
\begin{equation}\label{R}
\svfcrit=0.63, \qquad \mu_1=1, \qquad I_0=0.005, 
\end{equation}
and vary $p_1$ and $\Dar$ for $\mu_1$ and $\mu_2$. 

\begin{figure}[bt]
\centering
\includegraphics[width=0.32\linewidth,height=0.3\linewidth]{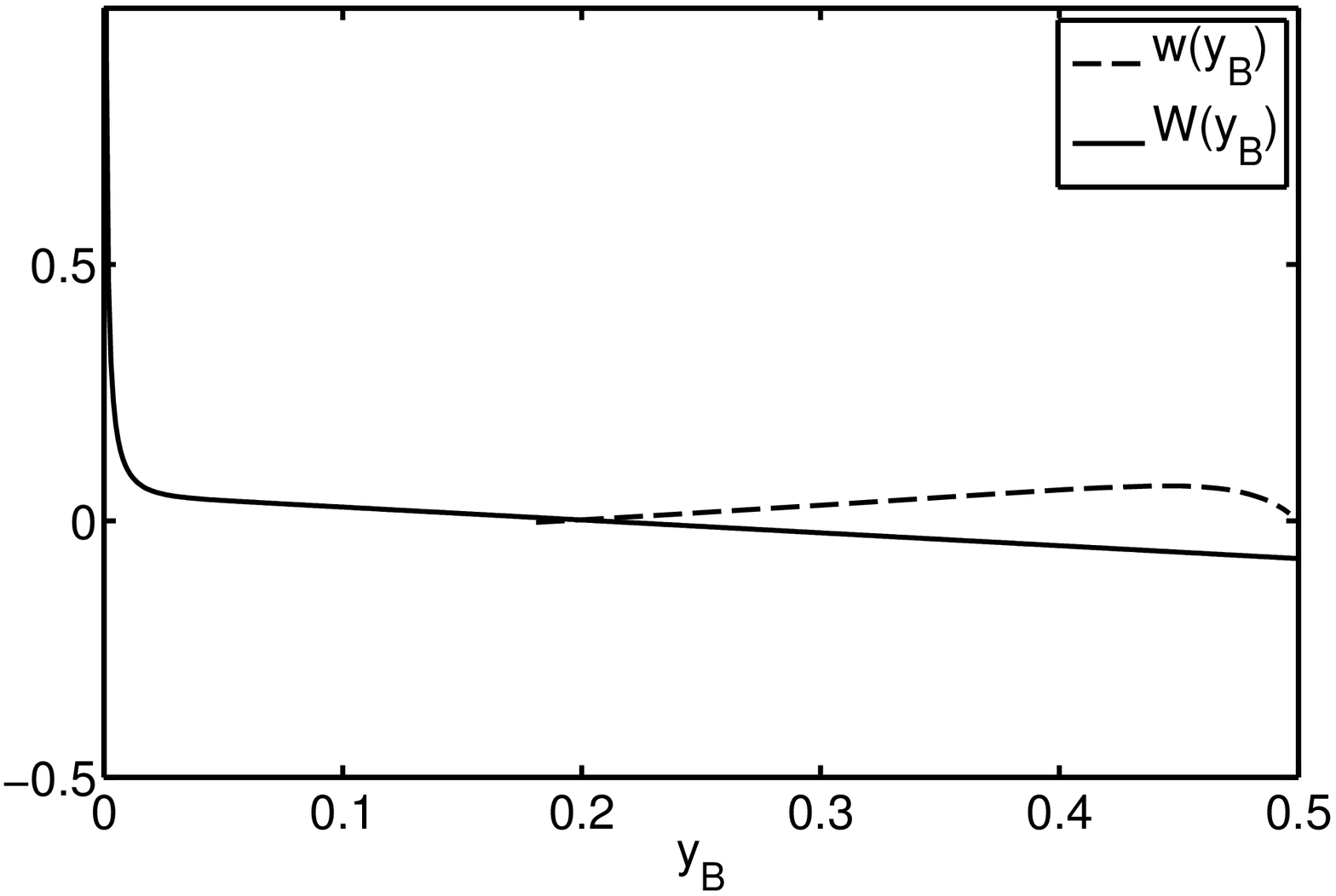}		
\includegraphics[width=0.32\linewidth,height=0.3\linewidth]{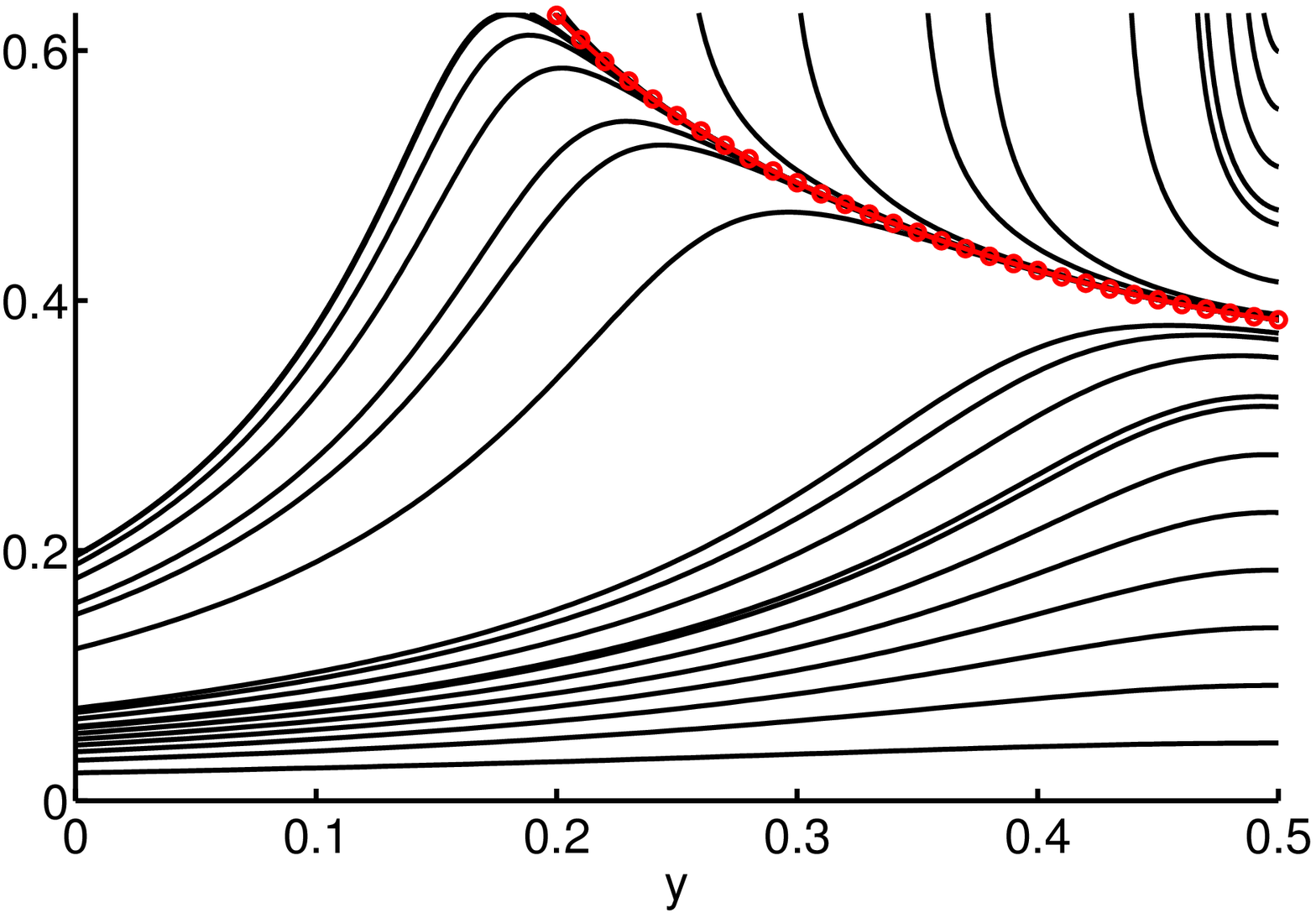}
\includegraphics[width=0.32\linewidth,height=0.3\linewidth]{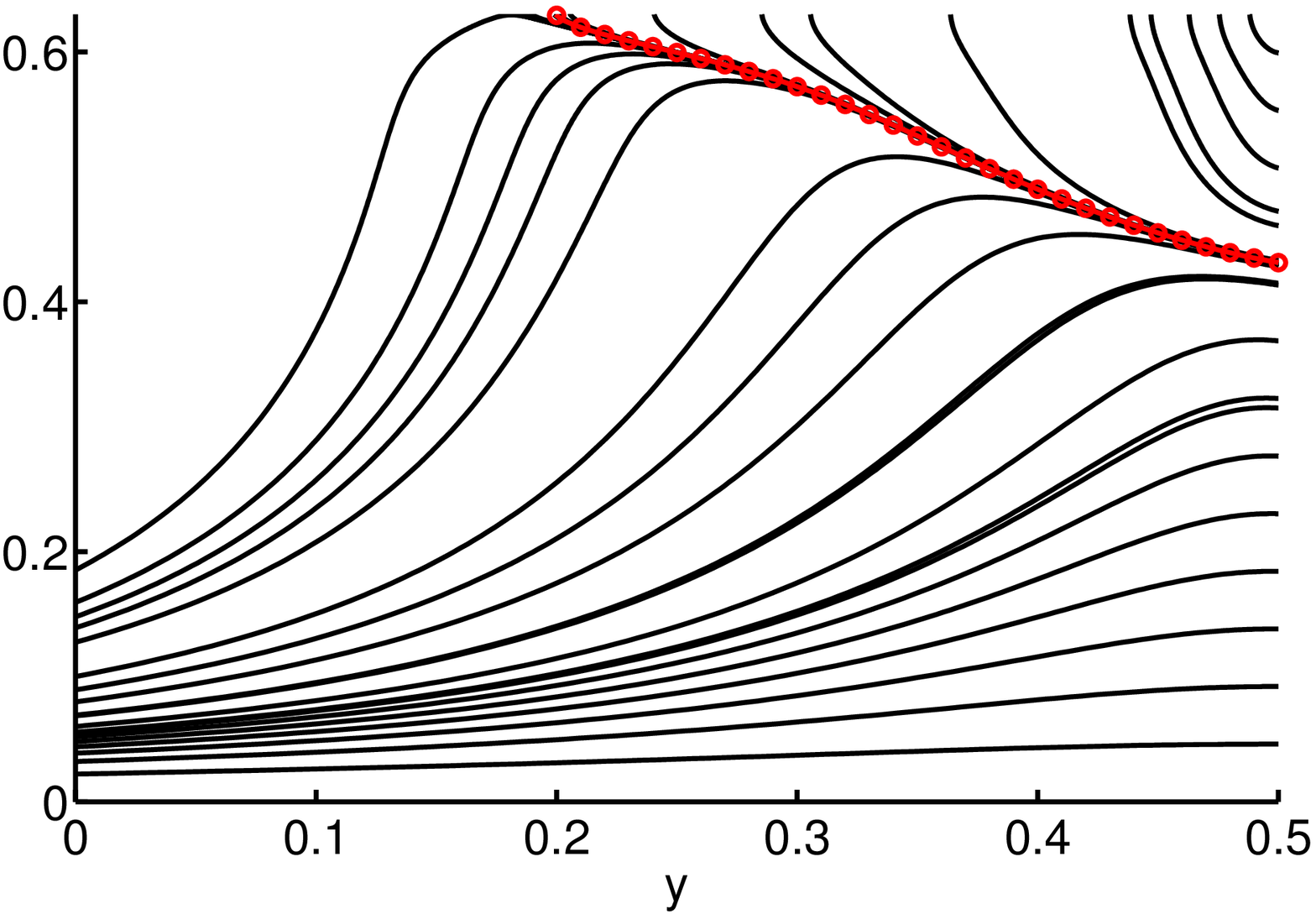}
\caption{
(left) The $w(y_B)$ solution curve together with the $W$-boundary condition and the projection of trajectories for \eqref{eqn:Ny_phi_w}, \eqref{eqn:wy_phi} onto the $\svf$-$y$-plane with initial values $w=0$, $\svf=\phi_0$ for a range of values. For the other parameters we used  $Da=1000$, $\mu_1=\mu_2=1$ (middle) and $\mu_1=1$, $\mu_2=1.5$ (right), where the latter case shows the impact of the viscosity term \eqref{muc_nondim} proposed in Boyer et al. \cite{Boyer2011}, which is zero for $\mu_1=\mu_2$.
}
\label{fig:phasespaceplot}
\end{figure}

For the solution of the boundary value problem \eqref{ODEformulation}, we proceed as follows. We solve \eqref{eqn:Ny_phi_w}, \eqref{eqn:wy_phi} as an initial value problem with initial values $\svf(1/2) = \phi_0$ and $w(1/2) = 0$ using e.g. Matlab's $ode15s$ solver. The solution is followed for decreasing $y$ until the volume fraction hits the value $\phi_{sc}$ or $y$ reaches zero. The situation is shown for a range of $\phi_0$ in figure~\ref{fig:phasespaceplot}. It turns out that typically there is a value $0 < \phi^* < \phi_{sc}$ so that the former case happens for $\phi_0\geq \phi^*$ and the latter if $\phi_0 < \phi^*$. We discard these values since only trajectories that intersect with $\svf = \phi_{sc}$ can lead to solutions of the boundary value
problem \eqref{ODEformulation}. For the remaining $\phi_0$ in the interval $[\phi^*, \phi_{sc}]$, we determine $y_B$ and $w(y_B)$ and plot the curve $(y_B(\phi_0),w(y_B(\phi_0)))$ as we vary $\phi_0$. The intersection of this curve with the graph of the function $W(y_B)$ (as defined in \eqref{freebc}), shown in figure~\ref{fig:phasespaceplot} identifies the unique value for $\phi_0$ that gives rise to a solution of \eqref{ODEformulation}. 
The corresponding trajectory is the unique symmetric solution with a single unyielded region and its projection on the $y$-$\svf$-plane in figure~\ref{fig:phasespaceplot} is emphasized by bullets. 

We note that the solution of the boundary value problem \eqref{ODEformulation}, 
can also be obtained by rewriting  \eqref{eqn:Ny_phi_w} for $w$, i.e.
\begin{align}
	w &= \frac{\left(\py N + \vf \, p_1\right)(1-\vf)}{\Dar \, \vf^2},
	\label{eqn:w_elimination}
\end{align}
and using it in \eqref{eqn:wy_phi} and in the boundary conditions. This yields an equation solely in $\vf$, i.e.
\begin{subequations}
\label{eqn:ODEsystemInPhi}
\begin{align}
		\py \left(\frac{\left(\py N + \vf \, p_1\right)(1-\vf)}{\Dar \, \vf^2}\right) &= \frac{p_1 y + N}{1- \vf} + \frac{1 }{\eta_n},
\end{align}
with boundary conditions
\begin{align}
		0 &= \py N + \svf \, p_1 & \text{ at } y &= \frac12, \\
		\vf &= \svfcrit & \text{ at } y &= y_B, \\
		\frac{\left(\py N + p_1\right)(1 - \svfcrit)}{\Dar \, \svfcrit^2} &= \frac{p_1 y_B + \mu_1}{\Dar^{\frac12} \svfcrit \tanh\left(\frac{\Dar^{\frac12} \svfcrit}{1 - \svfcrit} y_B \right)} & \text{ at } y &= y_B.
\end{align}
\end{subequations}
We transform the free-boundary problem \eqref{eqn:ODEsystemInPhi} 
into fixed-domain problem via
\begin{align}
	y = \left(y_B - \frac12\right) \zeta + \frac12,
\end{align}
where $\zeta \in [0,1]$, which introduces the free-boundary coordinate as an explicit parameter in the system and then we add the trivial differential equation for the constant $y_B$ to get the boundary value problem
\begin{subequations}
\label{eqn:ODEsystem_standardBVP}
\begin{align}
\frac{1}{y_B - \frac12} \partial_{\zeta} \left(\frac{\left(\frac{1}{y_B - \frac12} \partial_{\zeta} N + \vf \, p_1 \right)(1-\vf)}{\Dar\,\vf^2}\right) &= \frac{p_1 \left((y_B - \frac12)\zeta + \frac12\right) + N}{1 - \vf} + \frac{1}{\eta_n} \\
\partial_{\zeta} y_B &= 0
\end{align}
with boundary conditions
\begin{align}
	0 &= \partial_{\zeta} N + \left(y_B - \frac12 \right) \vf \, p_1 & \text{ at } \zeta &= 0 \\
	\vf &= \svfcrit & \text{ at } \zeta &= 1 \label{eqn:dirichlet_transformed} \\
	\partial_{\zeta} \vf &= -\frac{2(y_B - \frac12)}{5(1 -\svfcrit)} \frac{\Dar^{\frac12} \svfcrit (p_1 y_B + \mu_1)}{\tanh \left(\frac{\Dar^{\frac12} \svfcrit}{1 -\svfcrit} y_B\right)} + \frac25 \left(y_B - \frac12\right) p_1 & \text{ at } \zeta &= 1.
\end{align}
\end{subequations}

After solving for $\vf$, we can determine the remaining variables by first using \eqref{eqn:w_elimination} for $w$, next solving for $\fvf \fstresscmpt_{12}$ via \eqref{eqn:reduced_channel_ssm_with_yb:a} with $(\fvf \fstresscmpt_{12})(0) = 0$ and then get the fluid velocity via \eqref{eqn:reduced_channel_ssm_with_yb:d} with $u_f(1/2) = 0$. The solid variables are then easily computable by \eqref{eqn:reduced_channel_ssm_with_yb:b} and $u_s = u_f - w$.

\begin{figure}[tb]
\centering		
\includegraphics[width=0.6\linewidth]{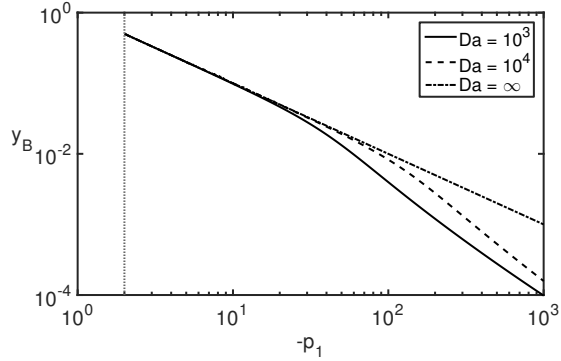}
\caption{
The dependence of the yield surface position $y_B$ on the pressure gradient magnitude $p_1$, for parameters \protect\eqref{R}, $\mu_2=1$. The solid curve shows the results for $\Dar = 1000$; the dashed curve for $\Dar = 10000$; the dotted-dashed curve for $\Dar = \infty$. The dotted vertical line represents the minimum pressure gradient $p_{\mathrm{min}} \approx 2$, where the unyielded region fills the entire channel.
}
\label{fig:p1_yb_plot}
\end{figure}

The dependence of the width of the unyielded zone on the pressure
gradient is shown in figure~\ref{fig:p1_yb_plot}.
As the magnitude of the pressure gradient $-p_1$ decreases, the thickness
of the unyielded zone increases until the interface between the yielded
and unyielded zone reaches the wall, that is, $y_B=1/2$, for $-p_1\leq p_\mathrm{min}$.
Setting $y=1/2$ in \eqref{freebc} and solving for $p_1$ gives an explicit expression
for this minimum pressure, 
\begin{equation}
 p_{\mathrm{min}} = \frac{\svfcrit \Dar \, \mu_1}{ \tanh \displaystyle\left(\frac{ \sqrt{\Dar}\, \svfcrit}{2(\svfcrit - 1)}\right)\,\sqrt{\Dar}\,(1-\svfcrit)^2+\Dar \,\svfcrit/2},
\end{equation}
below which the solid phase is jammed in the entire cross section of the channel.
On the other hand,  for $-p_1> p_{\mathrm{min}}$ 
the phase plane analysis suggests that there is always a unique and
strictly positive value for $y_B$ which moreover tends to zero as $p_1\to-\infty$.
We also note that for large $\Dar$ and fixed $p_1$,
the interface position tends to a finite value,
\begin{equation}
  y_B \to  -\frac{\mu_1}{p_1}. \label{eqn:yb_for_da_infinity}
\end{equation}

\paragraph{Remark}
For a finite length channel, the two free parameters $p_1$ and $p_c$ in the
solution for the solid fraction $\phi_s$ and for the velocity profiles $u_s$
and $u_f$ are typically fixed by e.g.\ inlet conditions for the solid and
liquid fluxes.  Notice that here we are reverting to the original parameter $p_1$ prior
to the rescaling in \eqref{pcrescaling1}.  Mass conservation dictates that for each phase,
the total fluxes must be constant along the channel, thus we have the conditions
\begin{subequations}\label{pcp1inl}
\begin{align}
2p_c \int_{0}^{1/2} (1-\phi(y;p_1/p_c)) u_f(y;p_1/p_c) dy &= \frac{1-\phi_{in} }{6}, \\
2p_c \int_{0}^{1/2} \svf(y;p_1/p_c) u_s(y;p_1/p_c) dy &= \frac{\phi_{in}u_{s,in}}{6}, 
\end{align}
\end{subequations} 
where $\phi_{in}$ and $u_{s,in}$ are the solid phase volume fraction and the scaling factor
for the parabolic solid phase velocity profile assumed at the inlet (see \eqref{inlet}).
Also recall that $u_{f,in}=1$ by our choice of the velocity scale. 
From our previous investigation, we know that the flow will always produce a
plug in the center where the solid phase is jammed, that is, $\phi=\phi_{sc}$.
This is possible by having a solid phase profile that is equal to $\phi_{sc}$
only in a very narrow region at the center of the channel and then rapidly decays to zero
towards the walls. Thus, the total solid flux through the cross section
can be arbitrarily small.  For such solutions, however, finite size effects may come
into play that are neglected in the continuum model.

\subsection{Asymptotic analysis for large Da}

\begin{figure*}[tb]
\centerline{
\includegraphics[height=4cm,width=0.33\linewidth]{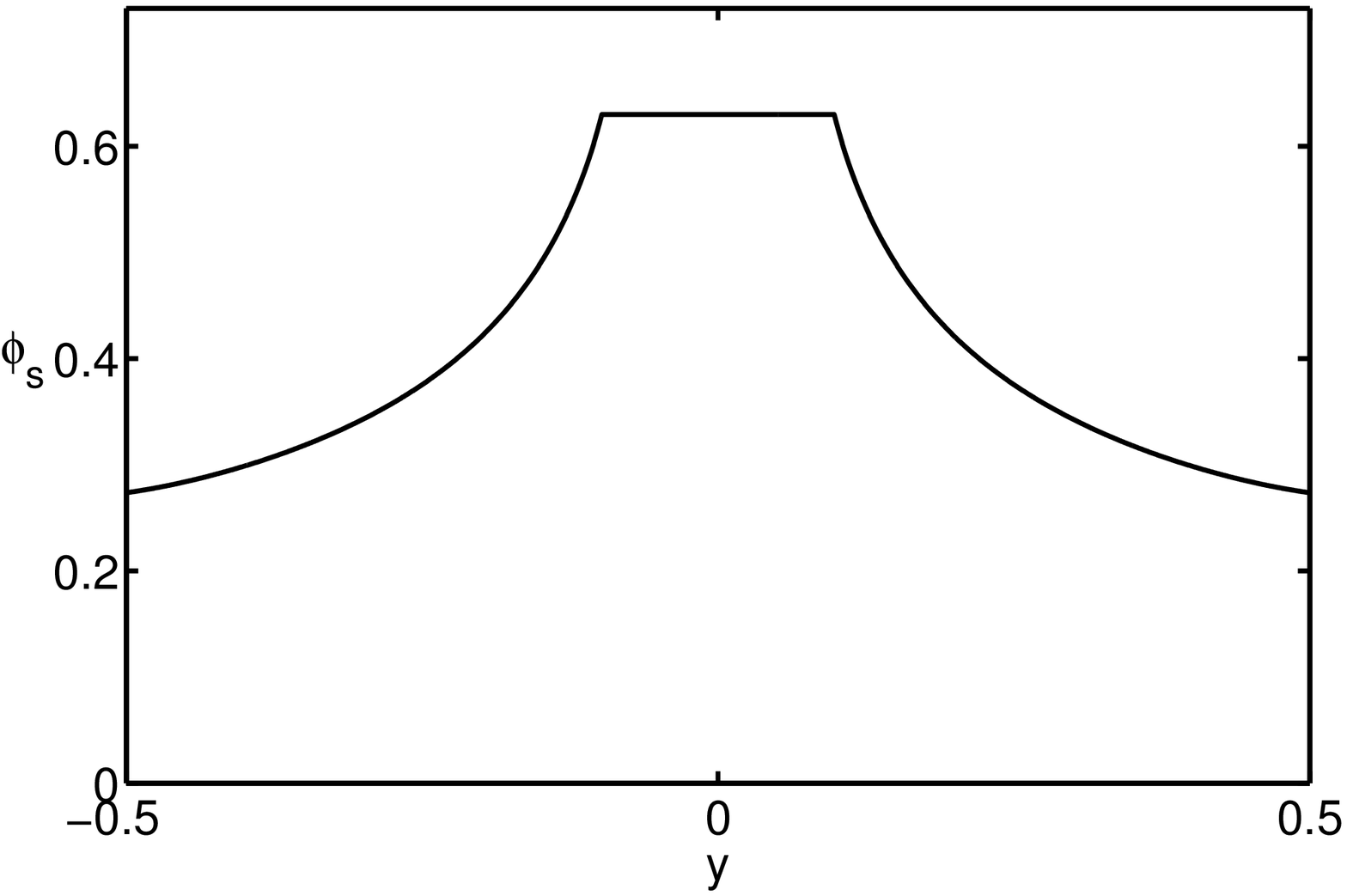}
\includegraphics[height=4cm,width=0.33\linewidth]{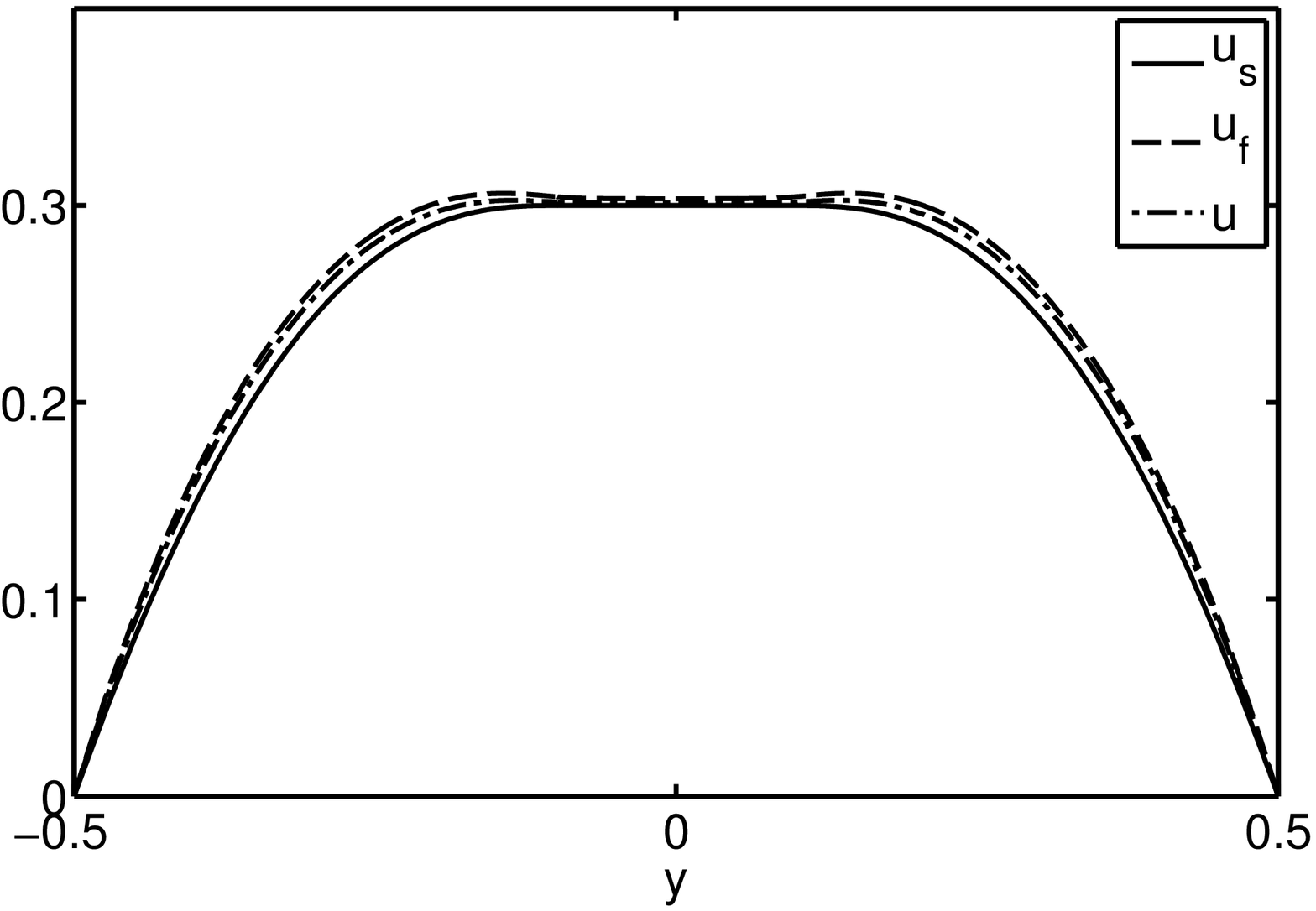}
\includegraphics[height=4cm,width=0.33\linewidth]{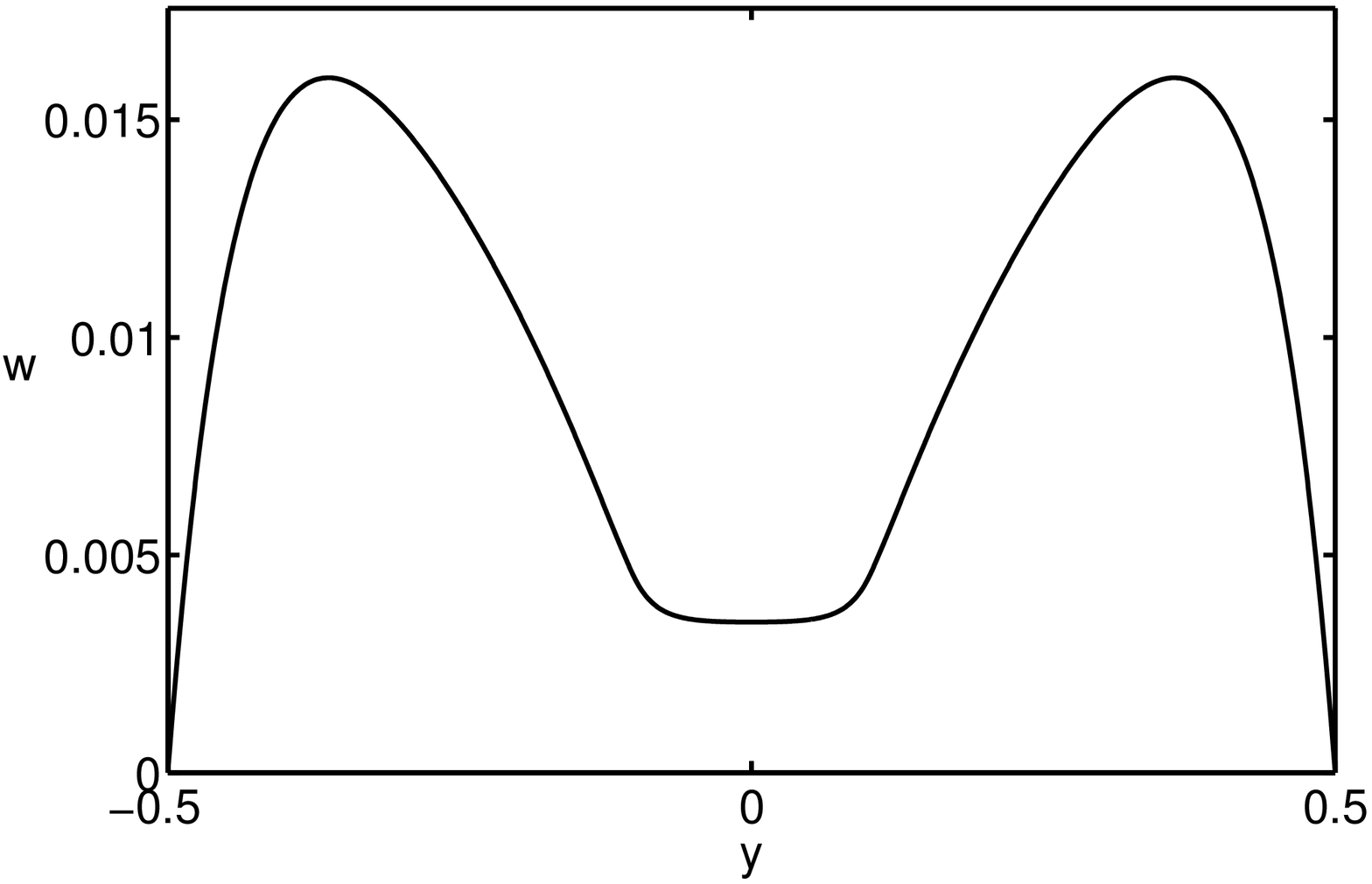}
}
\centerline{
\includegraphics[height=4cm,width=0.33\linewidth]{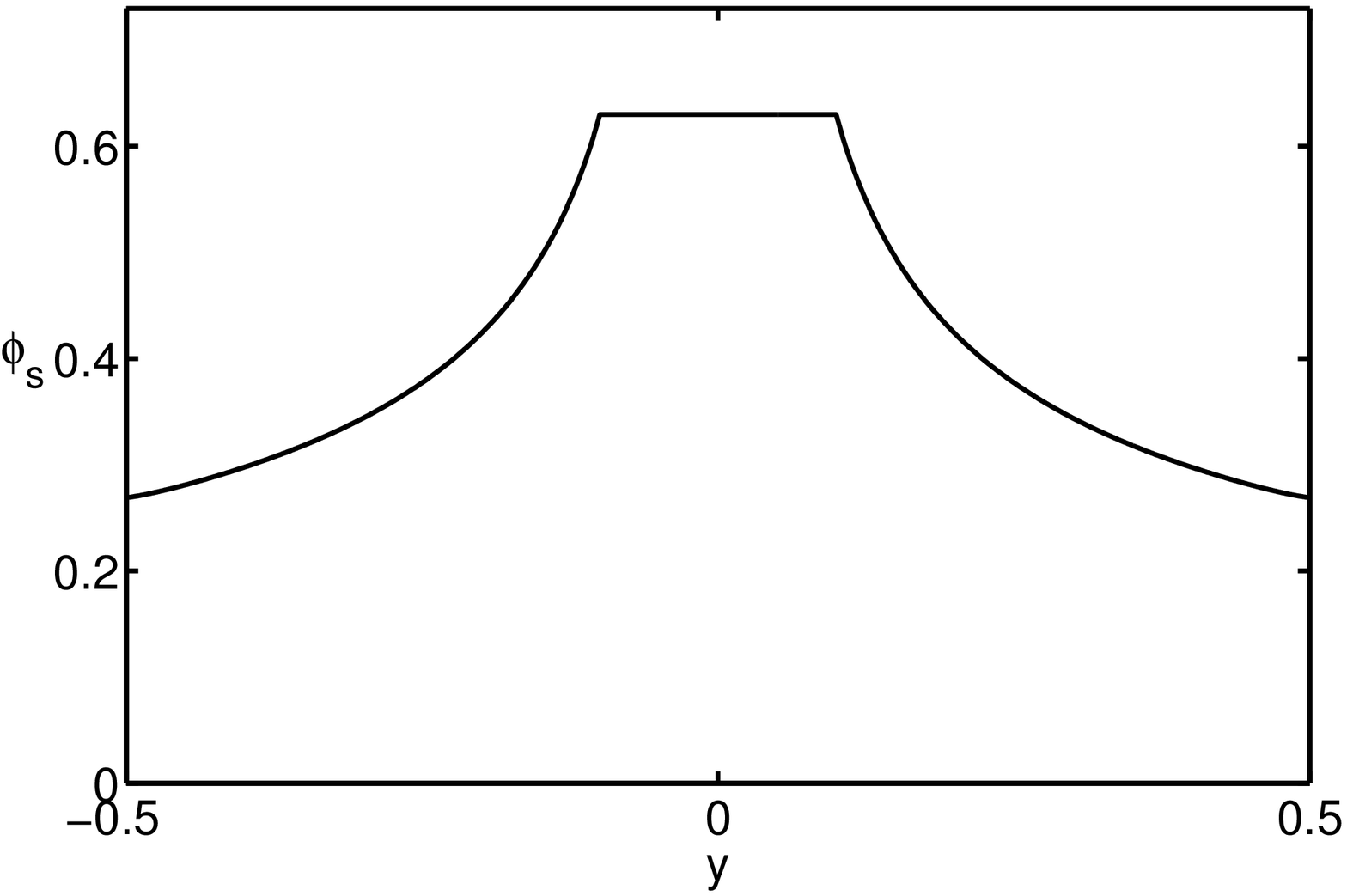}
\includegraphics[height=4cm,width=0.33\linewidth]{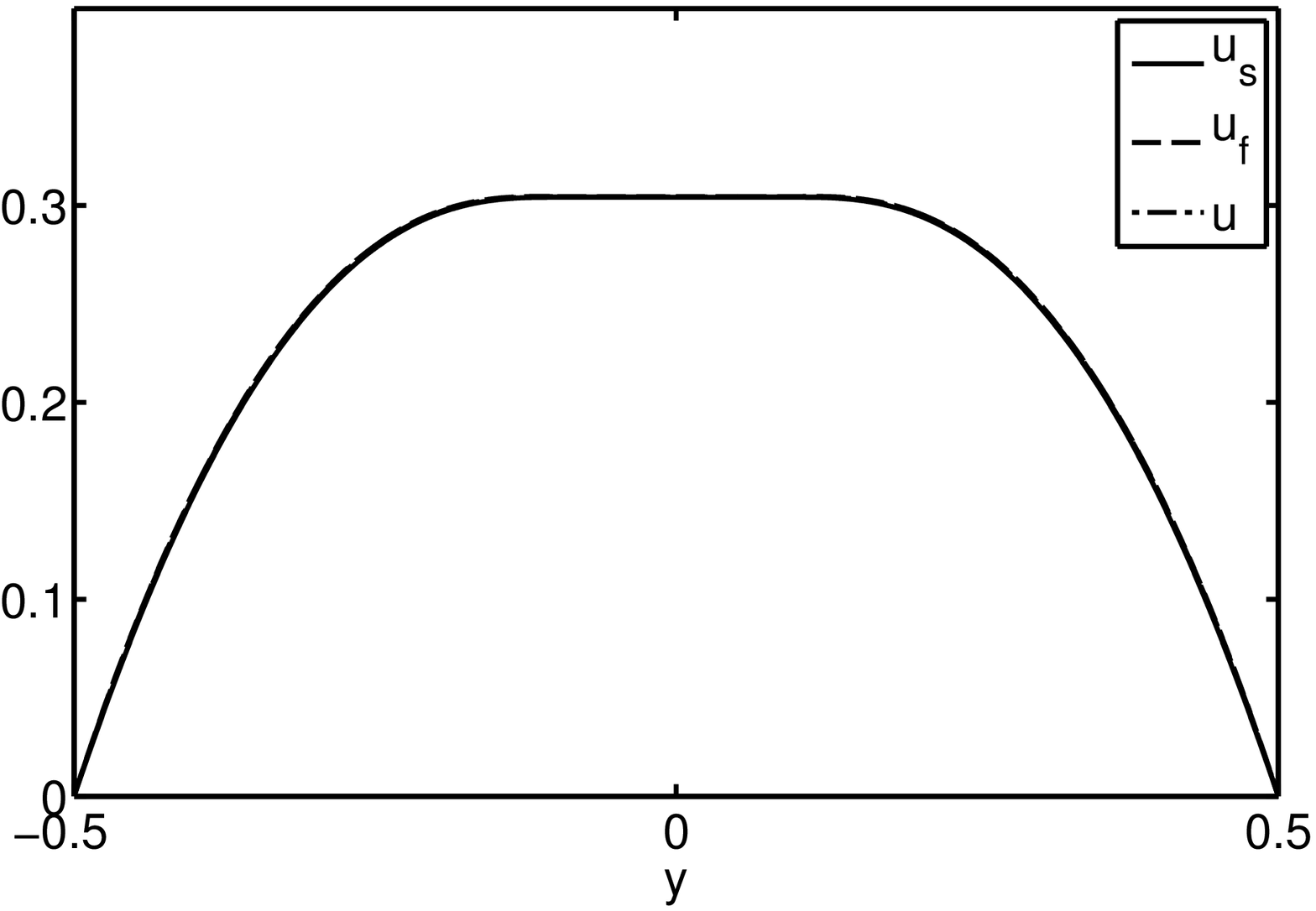}
\includegraphics[height=4cm,width=0.33\linewidth]{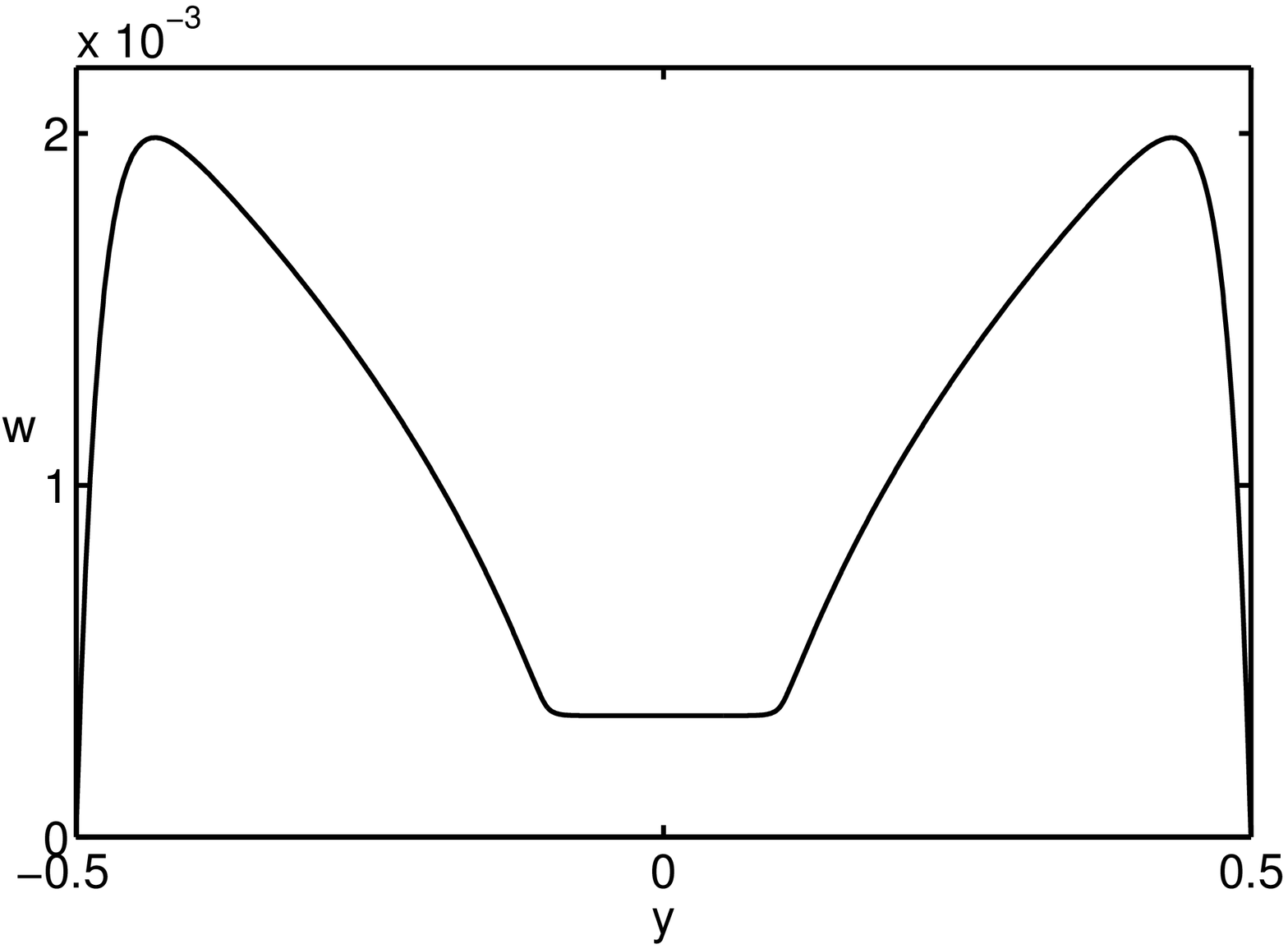}
}
\caption{
(left) The solid volume fraction $\svf$, (middle) the velocities $\svel, \fvel, \vel$ and (right) the velocity difference $w$ obtained by using the ODE problem \eqref{ODEformulation}. The parameters are given by \protect\eqref{R}, $\mu_2=1$, and $p_1=-10$. Top figures show results for $\Dar = 1000$ and bottom figures for  $\Dar = 10000$.
}
  \label{fig:solution_ode}
\end{figure*}

The stationary solutions show that the solid volume fraction increases
towards the channel center, where a finite size region at maximum
packing is located, see figure~\ref{fig:solution_ode}. The solid phase
velocity increases towards the center but is constant in the region
where $\phi=\svfcrit$, so that there, the solid phase is jammed, or
unyielded and effectively forms a porous medium.  The fluid velocity
increases at first but then decreases towards the center of the
channel, where it has a local minimum.  The difference between the two
velocities achieves a maximum away from the jammed region.  Moreover,
for growing $\Dar$ the solution of the stationary problem develops
boundary layers, in particular the velocity $w$ shows a pronounced
sharp drop as $y$ approaches the boundary $y=1/2$. In addition, as $\Dar$ increases,
the velocity difference $w$ decreases by approximately the same factor.
This observation and the large values of $\Dar$ suggest that we seek
and asymptotic approximation in the limit $\eps\equiv 1/\Dar\to 0$
with the ansatz $w=\eps \tilde w$. The asymptotic analysis of the
stationary solution will yield the key ideas for the derivation of a
new drift-flux model from the time-dependent two-phase flow model for
concentrated suspensions, which we then use to study the formation
and evolution of jammed regions in the flow.

For the remaining analysis we drop the tilde and obtain from \eqref{eqn:Ny_phi_w}
\begin{subequations}
\begin{equation}
w = p_1\frac{1-\vf}{\vf}+\frac{1-\vf}{\vf^2}N'(\vf)\py\vf . \label{asy_Ny_phi_w}
\end{equation}
Substitution the rescaled version of \eqref{eqn:wy_phi} yields a second order equation for $\vf$, 
\begin{equation}
\eps\,\py\left(p_1\frac{1-\vf}{\vf}+\frac{1-\vf}{\vf^2}N'(\vf)\py\vf\right)=\frac{p_1 y + N(\vf)}{1-\vf} + \frac{(\svfcrit - \vf)^2}{\vf^2}.  \label{asy_wy_phi}
\end{equation}
The boundary conditions at the yield interface $y=y_B$ are
\begin{align}
	\vf &= \svfcrit, \label{asy_bc_phib} \\
	\py\vf &= - \frac25\;\frac{\svfcrit}{1-\svfcrit}\;
	\frac{p_1 y_B+\mu_1}{\tanh\left(\frac{\svfcrit}{1-\svfcrit}{\eps^{-1/2}} y_B\right)} 
\,{\eps^{-1/2}}+ \frac25 p_1,
\end{align}
and at the channel wall we have $w=0$, so that
from \eqref{asy_Ny_phi_w}, we get
\begin{equation}
\py\vf= -p_1\,\frac{\vf}{N'(\vf)} \quad \text{ at } y = \frac12.
\end{equation}
\end{subequations}
Clearly, this is a singular perturbed problem with a boundary layer at $y=1/2$  and $y=y_B$. In fact, if we assume that $\vf$ and $y_B$ have asymptotic expansions 
\begin{equation}
\vf(y)=\vf_0(y) + \eps^{1/2}\vf_1(y) + O(\eps),\qquad y_B={y_B}_0 + \eps\,{y_B}_1 + O(\eps^2).
\end{equation}
Then to leading order we have 
\begin{equation}
0=\frac{p_1 y + N(\vf_0)}{1-\vf_0} + \frac{(\svfcrit - \vf_0)^2}{\vf_0^2}.    \label{statout}
\end{equation}
When we use this in \eqref{asy_Ny_phi_w} the boundary conditions for $w$ are not satisfied.

\subsubsection*{Boundary layer problem at $y=1/2$}

For the boundary layer variables $z=(\frac12 - y)\eps^{-1/2}$ and $\Phi(z)=\vf(y)$
the governing equation is 
\begin{subequations}
\be
\pz\left(-\eps^{1/2}\;p_1\frac{1-\Phi}{\Phi}+\frac{1-\Phi}{\Phi^2}N'(\Phi)\pz\Phi\right)=\frac{p_1/2 + N(\Phi) - \eps^{1/2}\;z}{1-\Phi} + \frac{(\svfcrit - \Phi)^2}{\Phi^2},   \label{asy_halfa}
\ee
with boundary condition at $z=0$
\be
(1-\Phi)\;N'(\Phi)\;\pz\Phi=\eps^{1/2}\; p_1\; (1-\Phi)\Phi . \label{asy_halfb}
\ee
\end{subequations}
Assume the asymptotic expansion of the inner variables can be written as
\be
\Phi(z)=\Phi_0(z) +\eps^{1/2}\Phi_1(z) + O(\eps), 
\ee
so that the solution satisfies to leading order the problem
\begin{subequations}
\begin{align}
\pz\left(\frac{1-\Phi_0}{\Phi_0^2}\, N'(\Phi_0)\,\pz\Phi_0\right) &= \frac{p_1/2 + N(\Phi_0)}{1-\Phi_0} + \frac{(\svfcrit - \Phi_0)^2}{\Phi_0^2}\\
\pz\Phi_0 &= 0 \qquad \mbox{at} \quad z=0^+
\end{align}
\end{subequations}
since $(1-\Phi_0)\;N'(\Phi_0)\neq 0$.
As $z\to\infty$ the solution approaches a constant, say $\Phi_0\to\Phi_{0,\infty}$, which satisfies
\be
\frac{p_1/2 + N(\Phi_{0,\infty})}{1-\Phi_{0,\infty}}+ \frac{(\svfcrit-\Phi_{0,\infty})^2}{\Phi_{0,\infty}^2}=0.
\ee
Hence, since for  $y\to (1/2)^-$ in the leading order outer problem, then 
\be\label{yhe}
0=\frac{p_1/2 + N\left(\vf_0(1/2)\right)}{1-\vf_0(1/2)} + \frac{\left(\svfcrit - \vf_0(1/2)\right)^2}{\vf_0^2(1/2)}.
\ee
Therefore, matching yields $\Phi_{0,\infty}=\vf_0(1/2)$. 

It is straightforward to solve the next order problem to obtain
\be\label{Phi1sol}
\Phi_1(z)= \frac{A_2 N'(\ph)-p_1\ph A_1}{A_1^{3/2}N'(\ph)}\exp\left(-\sqrt{A_1} z\right)  + \frac{A_2}{A_1} z,
\ee 
where
\begin{subequations}
\begin{align}
A_1 &=  \frac{\pht}{(1-\ph)^2} + \frac{\pht}{N'(\ph)} \frac{\frac12 p_1 + N(\ph)}{(1-\ph)^3} 
\notag\\&\qquad\qquad\qquad\qquad - \frac{2}{N'(\ph)} \frac{\svfcrit}{\ph} \frac{\svfcrit-\ph}{1-\ph},\qquad\qquad\\
A_2 &= \frac{p_1}{N'(\ph)} \frac{\pht}{(1-\ph)^2},
\end{align}
\end{subequations}
thus, using \eqref{yhe}
\begin{align}
\frac{A_2}{A_1}
&=
p_1\left[
{N'(\ph)}
+ \frac{\left(\svfcrit - \ph\right)\left(\pht-2\svfcrit+\svfcrit \ph\right)}{\phc}
\right]^{-1} .
\end{align}
Taking the $y$-derivative of \eqref{statout} we get
\begin{equation}\label{dyphi0}
\partial_y \phi_0=-p_1\left[N'(\phi_0)+\frac{\svfcrit-\phi_0}{\phi_0^3}
\left(\phi_0^2-2\svfcrit+\phi_0\svfcrit\right)\right]^{-1}.
\end{equation} 
Therefore, the linear term in the expansion of the outer solution $\phi_0$ and
in the inner solution $\Phi_1$, see \eqref{Phi1sol}, match as required.

\subsubsection*{Boundary layer problem at $y=y_B$}
Similarly, we let the boundary layer variables be 
\begin{equation}
	\xi=\frac{y-y_B}{\eps^{1/2}}, \qquad \varphi(\xi)=\vf(y).
\end{equation}
To leading order the problem now reads
\begin{subequations}
\be
\pxi\left(\frac{1-\varphi_0}{\varphi_0^2}\, N'(\varphi_0)\,\pxi\varphi_0\right)= \frac{ p_1 \; {y_B}_0+ N(\varphi_0)}{1-\varphi_0} + \frac{(\svfcrit - \varphi_0)^2}{\varphi_0^2},
\ee
with boundary condition at $\xi=0^+$
\be
\varphi_0(0)=\svfcrit
\ee
and
\be
\pxi\varphi_0(0)=-\frac25\; \frac{\svfcrit}{1-\svfcrit}\; (p_1 \; {y_B}_0+ \mu_1)=0.
\ee
\end{subequations}
Note, if we assume that in the leading order outer equation, $\vf$ also satisfies $\vf=\svfcrit$ at $y=y_B$ then we must have that $p_1\; {y_B}_0 + \mu_1 = 0$, since $N(\svfcrit)=\mu_1$. Hence, the second boundary condition is also zero. 
This suggests $\varphi_0=\svfcrit$. Matching this to the leading order outer problem
\be\label{matchii}
\frac{p_1\;{y_B}_0 + N(\vf_0({y_B}_0)}{1-\vf_0(y_B)} + \frac{(\svfcrit-\vf_0(y_B))^2}{\vf_0^2(y_B)} = 0.
\ee
Hence, $\vf_0(y_B)=\svfcrit$. Solving the next order problem 
\begin{subequations}
\be
\pxixi \varphi_1= \left(\varphi_1 - \frac25\, p_1\,\xi\right)\frac{\svfcrit^2}{(1-\svfcrit)^2}
\ee
with boundary conditions 
\be
\varphi_1(0)=0, \quad
\pxi\varphi_1(0)=\frac25\;p_1
\ee
\end{subequations}
gives
\be
\varphi_1(\xi)= 
\frac25 p_1 \xi.
\ee
This needs to be matched with the linear term in the Taylor expansion of the
leading order outer solution $\phi_0$, which can be obtained by taking the limit 
$\phi\to\svfcrit$ in \eqref{dyphi0}. That limit gives $\partial_y\phi_0(y_B)=-p_1/N'(\svfcrit)
=-p_1/(-5/2)$, that is, the coefficients are equal, hence the terms match.

%

\section{Drift-flux model for plane Poiseuille flow}
While drift-flux models have been proposed to study the evolution of two-phase flows of suspensions \cite{Leighton1987,Phillips1992} and are also used as transport equations for a suspended phase and combined with hydrodynamic equations \cite{Cook2008,Murisic2013} a systematic asymptotic derivation from the underlying two-phase model is still open. 
Here, we will use matched asymptotics along the lines of the analysis of the stationary problem, for the derivation of a new drift-flux model for the cross-sectional flow of the channel. Our analysis shows that the inclusion of the boundary layers leads to a drift-flux model that naturally accounts for the  shear-induced flux of the suspended phase away from the boundaries. 
Moreover, the constitutive law for concentrated suspensions leads to the appearance of unyielded and yielded regions, which needs to be captured by the new  drift-flux model. 

\subsection{Asymptotic derivation of the drift-flux model}\label{subsec:ad}
To capture the evolution towards a Bingham-type flow it will be instructive to investigate the problem for the cross-section. We assume therefore that all the variables depend only on $y$ and $t$, except for the pressure variables, which also depend on $x$.

As in our previous section, the drift-flux regime is established for large $\Dar$ and small velocity differences $w$, and in addition on a long time scale. Hence, we let $\eps=1/\Dar$
and
\be
 w_1=\eps w_1^*,\quad w_2=\eps w_2^*, \quad t=\frac{t^*}{\eps}.
\ee
The governing equations are then, after we drop the ``$*$'' 
\begin{subequations}
\begin{align}
\pt\phi - \py(\phi\, (1-\phi)\, w_2) &= 0\quad \\
-\py\left[(1-\phi)\,\py v_1 + \eps (1-\phi)\,\py(\phi w_1)\right] +(1-\phi)\px p_f &= -\frac{\phi^2}{1-\phi} w_1\\
-\py\left[2\eps (1-\phi)\,\py(\phi w_2)\right] + (1-\phi)\py p_f &= -\frac{\phi^2}{1-\phi} w_2 \\
-\py\left[\phi\eta_s\py v_1 - \eps\phi\eta_s\py((1-\phi) w_1)\right] + \phi\px \fprs &= \frac{\phi^2}{1-\phi} w_1\\
\py\left[2\eps\phi\,\py((1-\phi) w_2)\right] + \svf \py \fprs + \py \prs_c &= \frac{\phi^2}{1-\phi} w_2\\
\hspace*{0cm}\prs_c=\eta_n(\phi)\left[(\py v_1 - \eps\py((1-\phi) w_1))^2 + 2\eps[\py((1-\phi) w_2)]^2\right]^{1/2} &+ \eps^4 \svf 
\end{align}
and no-slip conditions at $y=\pm 1/2$
\be
\hspace*{6cm}v_1=0,\quad w_1=0,\quad w_2=0.
\ee
\end{subequations}

To leading order we obtain for the outer problem 
\begin{subequations}
\begin{align}
\pt\phi - \py(\phi(1-\phi)w_2) &= 0 \label{sysa}\\
-\py[(1-\phi)\py v_1] + (1-\phi)\px p_f &= -\frac{\phi^2}{1-\phi} w_1 \label{sysb}\\
(1-\phi) \py p_f &= -\frac{\phi^2}{1-\phi} w_2 \label{sysc}\\
-\py[\phi\eta_s\py v_1] + 
\phi \px p_f + \px p_c
&= \frac{\phi^2}{1-\phi} w_1 \label{sysd}\\
\phi \py p_f + \py p_c
&= \frac{\phi^2}{1-\phi}w_2 \label{syse}\\
p_c &= \eta_n\left|\py v_1\right|, \label{sysf}
\end{align}
and no-slip conditions at $y=\pm 1/2$
\be
v_1=0,\quad w_1=0,\quad w_2=0.
\ee
\end{subequations}
We note that for ease of notation we have  dropped the indices in the variables that denote the leading order solutions. 
Adding \rf{sysc} and \rf{syse} yields $\py(p_f + p_c) = 0$, hence $p_f+p_c=f(x)$. Adding \rf{sysb} and \rf{sysd} yields
\be
-\py\left(\left[\phi\eta_s + (1-\phi)\right]\py v_1\right) + \px (p_f + p_c) = 0. 
\ee
Since the left hand side is only dependent on $y$ and the right hand side only on $x$, they must be constants. Thus, defining $\px(p_f + p_c) = p_2$, so that after integration
\be
\left[\phi\eta_s + (1-\phi)\right]\py v_1 = p_2 y + \alpha. \label{systh}
\ee
Adding $(1-\phi) \py p_c$ on both sides of \eqref{sysc} yields
\be
\py p_c = \frac{\phi^2}{(1-\phi)^2}w_2.
\ee
We have
\be
w_2 = \frac{(1-\phi)^2}{\phi^2}\py\left(\eta_n|\py v_1|\right)=\frac{(1-\phi)^2}{\phi^2}\py\left(\eta_n\dot\gamma\right).
\ee
In addition note that from \rf{systh} we obtain 
\be
\py v_1 = \frac{p_2\, y}{\phi\eta_s +1-\phi},
\ee
where due to symmetry we have set $\alpha=0$. Since $p_2<0$ the negative of this expression will always be positive and we set 
\be
\dot\gamma=-\frac{p_2\, y}{\phi\eta_s +1-\phi},
\ee
so that
\be\label{w2}
w_2=-p_2\frac{(1-\phi)^2}{\phi^2}\partial_y 
\left[\frac{\eta_n y}{\phi \eta_s + 1-\phi}\right].
\ee
Hence, we obtain for the drift-flux model 
\be
\pt\phi = -p_2\py\left[\frac{(1-\phi)^3}{\phi} \py\left(\frac{y}{N(\phi)+\frac{1-\phi}{\eta_n(\phi)}}\right)\right].
\label{driftfluxout}
\ee

We note at this point that the drift-flux model we have just derived \rf{driftfluxout} is a nonlinear diffusion equation which admits constant solutions, say $\phi_0$. Linearizing about these base states by making the ansatz $\phi(t,y)=\phi_0+\delta\,\phi_1(t,y)+O(\delta^2)$ we obtain to $O(\delta)$
\be
\pt \phi_1= -p_2\,\py\left[\frac{M'(\phi_0)}{F(\phi_0)} \phi_1 - M(\phi_0)\frac{F'(\phi_0)}{F^2(\phi_0)} \py\left(y\phi_1\right)\right],
\ee
where $M(\phi)={(1-\phi)^3}/{\phi}$ and $F(\phi)=N(\phi)+ (1-\phi)/\eta_n(\phi)$. 
We supplement this equation with boundary conditions and assume no-flux conditions at the wall $y=1/2$. Indeed, as shown in appendix \ref{boundary_layer_appendix}, matching to a boundary layer there gives $w_2=0$.
We seek solutions that are symmetric with respect to the middle axis of the channel, thus we also impose $w_2=0$ at $y=0$. 

\begin{figure}[bt]
	\centerline{
		\includegraphics[width=0.7\linewidth]{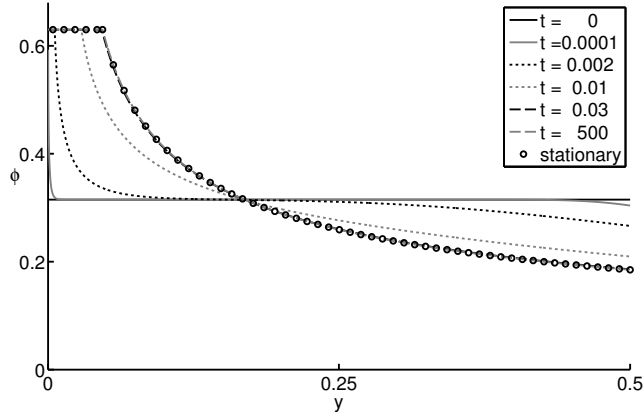}
	}
	\caption{Time evolution of solid volume fraction using the outer drift-flux approximation \eqref{driftfluxout}
for the parameters from \eqref{R} with $\mu_1 = \mu_2$ and $p_1 = -10$, starting from an initial uniform profile of $\vf(0,y) =  \svfcrit/2$.
The profile first changes near the channel center and wall. 
The volume fraction increases near the center until maximum packing is reached, which spawns an unyielded region. This unyielded region then grows so that $y_B$ approaches the stationary value obtained from \eqref{eqn:yb_for_da_infinity} and $y$ converges to the stationary solution.
}
	\label{fig:driftfluxouter_asderived}
\end{figure}

Clearly, if $F'(\phi_0)<0$,
which holds true for all $\phi_0 \in [0,\svfcrit]$ as long as $\mu_2 \ge \mu_1$ and $I_0 \ge 0$,
then any perturbation of the constant base states is damped out and the flow remains. 
But we note that non-zero constant solutions of \eqref{driftfluxout} do not satisfy the boundary conditions. Hence, we expect a nonlinear structure to arise from the interplay between the drift-flux equation and the no-flux condition.
Indeed, the flux of the solid phase leads to an increase of $\phi$ at the center of the channel,
until the solid volume fraction reaches $\phi_{sc}$ there and jamming of the solid phase 
occurs in a region $y<y_B$ with a time-dependent free boundary $y_B(t)$. In fact, $w_2=0$ cannot be achieved by the channel center $y=0$
at the right hand side of \eqref{w2}, thus the jammed region emerges instantaneously, that is,
$y_B(t)>0$ for all $t>0$. In the jammed region, the solid volume fraction is constant so that mass conservation gives $w_2=w_2(t)$. Assuming symmetry at $y=0$ then fixes $w_2$ to be zero to all orders in $\eps$ for $0<y<y_B$. At $y=y_B$, we therefore impose $\phi=\phi_{sc}$ and $w_2=0$, so that we have two boundary conditions as required at a free boundary.
Fig.~\ref{fig:driftfluxouter_asderived} shows a numerical solution for the drift-flux model \eqref{driftfluxout} with 
\begin{align}
 \py\left(\frac{y}{N(\phi)+\frac{1-\phi}{\eta_n(\phi)}}\right) &= 0
\label{eqn:no_flux_bc_drift_flux}
\end{align}
imposed at $y = 1/2 $ and at $y=y_B$.
The second condition 
\begin{align}
	\vf &= \svfcrit \quad \text{ at } y = y_B
\end{align}
is used to update the free boundary $y_B$. A central finite difference scheme of second order with a fully implicit Euler-Euler-2-step method was used to discretize the problem. 
The results in fig.~\ref{fig:driftfluxouter_asderived} clearly display 
the emergence and evolution of the jammed region in the cross-sectional channel flow.

The evolution eventually tends to a stationary state that can be obtained from
\eqref{driftfluxout} by letting $\pt\phi = 0$. Integrating once and using 
\eqref{eqn:no_flux_bc_drift_flux} at $y=1/2$ and then integrating once again leads to 
\be
\frac{y}{N(\phi)+\frac{1-\phi}{\eta_n(\phi)}}=c_1.
\ee
With $c_1=-1/p_1$ we recover the stationary outer equation \eqref{statout}
from section~\ref{sec:channel_flow}. Its value is fixed here by the requirement that the total amount of solid phase material
\begin{equation}
V_s=\int_{0}^{1/2} \phi \; \ud y
\end{equation}
is equal to the total amount present in the initial condition $\phi(y,0)$ for the time-dependent problem. This follows from the observation that $V_s$ is a conserved quantity for the time-dependent problem.
The corresponding solution is indicated
in the figure by open circles. It agrees well with the long time profile for $\phi$
obtained from the time-dependent problem.


\paragraph{Remark} \label{app:hyperbolicity}
Before we continue with our analysis we like to note that it is well-known that the non-viscous one-pressure two-fluid system contains an ill-posedness, that manifests itself in the occurrence of complex characteristics in the system and a subsequent loss of hyperbolicity in time, see also the recent discussion in  \cite{Lhuillier2013}. The problem exists even for models that include viscous terms \cite{Prosperetti1987}.
Although there has been progress towards a mathematical understanding during the last decade, see e.g.\  \cite{Keyfitz2003541} and references therein, 
until now there is no consistent and at the same time physically meaningful approach that resolves this problem. For the model above, we propose a mathematically motivated regularization to avoid the problem of loss of hyperbolicity. If we introduce a modified 
expression for the collision pressure of the form
\begin{align}
	p_c = \eta_n(\svf) |\ssrt| + c \svf,
\end{align}
where $c$ is a constant regularization parameter
that is only slightly larger than 
\begin{align}
	c_m= \frac{(u_f-u_s)^2}{\Dar^2},
\end{align}
then the
equation is hyperbolic i.e.\ all characteristics are real \cite{Lhuillier2013},
but the additional term does not interfere 
with the derivation of the drift-flux model 
for $\Dar\gg 1$.


\section{Conclusion and Outlook}
\label{sec:conclusion}

In this study we systematically derived a new two-phase model through ensemble averaging along the lines of Drew et al.~\cite{Drew1971} while incorporating recent non-Brownian constitutive laws by Boyer et al.~\cite{Boyer2011} for the shear and normal viscosities for concentrated suspensions. 

Our study of plane Poiseuille flow using the two-phase model shows the existence of unyielded or jammed regions. The width of such a region depends on the value of the applied pressure for given volume fraction of the solid phase. We also demonstrated the dependence of the profile for the volume fraction $\svf$ on the so-called ``viscous number'', which can induce a qualitative change in the approach towards maximum volume fraction.
$\Dar$ is typically very large because of small particle sizes, and for these values $w_1 = u_f - u_s$, i.e. the difference between the solid and liquid phase velocity, develops a boundary layer at the channel walls and at the interface between unyielded and yielded regions.

Using matched asymptotic expansions, we then derive a new drift-flux model that 
allows for the emergence of jammed regions. 
Our asymptotic analysis shows that in order for the drift-flux model to correctly capture shear-induced particle migration the boundary layer structure of the solution has to be resolved and matched to the ``outer'' problem of the drift-flux model. 

Our numerical solutions of the drift-flux model reveal how the
jammed region emerges first at the center and then expands until the
stationary state is reached.  It would be interesting to relate this
evolution to experimental results on the transition length over which
a steady state develops in space from homogeneous inlet conditions.

Our analysis suggests that the boundary layer acts as a source for the particle migration towards the unyielded region. The quantities $w_1$, $w_2$, which denote the difference between the velocities $u_f$, $u_s$ and $v_f$, $v_s$, respectively, are by $O(\eps)$ smaller than the actual flow variables. The fact that the particle transport acts on a different time scale than the phase-averaged flow field also indicates how to systematically develop an asymptotic theory leading to a complete coupled flow model that includes both transport and jamming of particles. Such an analysis could also rationalize some suspension flow models that are found in the literature.
In fact, the methods presented in this study should also enable the systematic derivation of drift-flux models for more complex flow geometries, for example at the free boundary between the suspension and the surrounding atmosphere, and will be  part of our future work. 


\section*{Acknowledgements}
The authors are extremely thankful to Prof. Andrew Fowler (Mathematical Institute, University of Oxford) for very fruitful discussions.

AM is grateful for the support by KAUST (Award Number KUK-C1-013-04). TA and BW gratefully acknowledges the support by the Federal Ministry of Education (BMBF) and the state government of Berlin (SENBWF) in the framework of the program {\em Spitzenforschung und Innovation in den Neuen L\"andern} (Grant Number 03IS2151)

\begin{appendix}

\section{Derivation of the two-phase flow model}\label{app:derivation}

\subsection{Averaging rules}\label{app:prelim}

We will follow the mathematical framework by Drew and Passman \cite{Drew1983},\cite{Drew1999} in this section.
Let $f$ and $g$ be arbitrary measurable functions, $c$ a constant and $\ensavg{\cdot}$ an average operator obeying 
the so-called Reynolds' rules
\begin{align}
 \ensavg{f + g} &= \ensavg{f} + \ensavg{g} \label{eqn:reynolds_linearity} \\
 \ensavg{\ensavg{f}g} &= \ensavg{f}\ensavg{g} \label{eqn:reynolds_product_multiple_average} \\
 \ensavg{c} &= c \label{eqn:reynolds_constant},
\end{align}
the Leibniz' rule
\begin{equation}
 \ensavg{\partial_t f} = \partial_t \ensavg{f}
\end{equation}
and the Gauss' rule 
\begin{equation}
 \ensavg{\partial_i f} = \partial_i \ensavg{f} \label{eqn:GaussRule}.
\end{equation}
The functions should be weakly differentiable up to the required order. Admissible operators are for example the
volume average \cite{Whitaker1986}, \cite{Kolev2005}, time averages \cite{Ishii2011}, the
ensemble average \cite{Drew1999} or a mixture of these \cite{Drew1971}. However, note the
derivatives are defined in the sense of distributions in this work. This implies $\ensavg{\nabla f}$ can have a
Dirac delta property yielding additional surface integrals, whereas in classical theories the
Leibniz' and Gauss' rule are written explicitly with surface integrals, c.f. \cite{Drew1999} and \cite{Whitaker1986}.

We further need a component indicator function
\begin{equation}
 X_k(\vc{x},t) =
\begin{cases}
 1, &\text{ if } (\vc{x},t) \in K \\
 0, &\text{ if } (\vc{x},t) \not\in K
\end{cases}
\end{equation}
with $K$ the set of states of the $k$-th-phase.
In our model we use the average operator in a weighted form. There are in general two averages in use, the intrinsic or phasic average
\begin{equation}
 \pavg{g} \equiv \frac{\ensavg{X_k g}}{\ensavg{X_k}}
\end{equation}
and the mass-weighted or Favr\'e average (in its three common forms)
\begin{equation}
 \favg g \equiv \frac{\pavg{\rho g}}{\pavg{\rho}} = \frac{\ensavg{X_k \rho g}}{\ensavg{X_k} \frac{\ensavg{X_k
\rho}}{\ensavg{X_k}}} = \frac{\ensavg{X_k \rho g}}{\ensavg{X_k \rho}}.
\end{equation} 
When we have multiple indicator functions, an index states the indicator function we used in the average, e.g. $\pavg{g}_s$ means we used $X_s$ in the average. We define a fluctuation field (cf. \cite{Drew1999}) as
\begin{align}
 \pder g &:= g - \pavg{g} \\
 \fder g &:= g - \favg{g}
\end{align}
and due to the Reynolds rules $\pavg{\pder g} = \favg{\fder g} = 0$ holds.
This splitting together with the Reynolds rules yields the identity
\begin{equation}
 \pavg{f g} = \pavg{f} \pavg{g} + \pavg{\pder f \pder g} \label{eqn:phasic_product_rule}
\end{equation}
and similar for the Favr\'e average
\begin{equation}
 \favg{f g} = \favg{f} \favg{g} + \favg{\fder f \fder g} \label{eqn:favre_product_rule}.
\end{equation}

The characteristic function fulfills the so-called topological equation (cf. \cite{Drew1999})
\begin{equation}
 \partial_t X_k + \vc{u}_i \cdot \nabla X_k = 0
\end{equation}
with $\vc{u}_i$ the interface velocity.

\subsection{The two-phase flow model}\label{app:twophase}

Multiplication of \eqref{eqn:balance}-\eqref{eqn:jump_cond_def2} with $X_k$, followed by usage of the average operator and its linearity together with Gauss' and Leibniz' rules yield
\begin{align}
\label{pa1}
 \partial_t \ensavg{X_k \rho} + \nabla \cdot \ensavg{X_k \rho \vel} &= \ensavg{\rho(\partial_t X_k + \vel_i \nabla \cdot X_k)} \\
 &\qquad + \ensavg{\rho (\vel-\vel_i) \cdot \nabla X_k} \\
 \partial_t \ensavg{X_k \rho \vel} + \nabla \cdot \ensavg{X_k \rho \vel \otimes &\vel} 
 - \nabla \cdot \ensavg{X_k \ts{T}} = \ensavg{X_k \vc{f}} \\ 
 \label{pa2}
 + \ensavg{(\partial_t X_k + \vel_i \cdot \nabla X_k) &\rho \vel} \\
 + \ensavg{[(\vel - \vel_i) \cdot \nabla &X_k] \rho \vel} - \ensavg{\nabla X_k \cdot \ts{T}}.
\end{align}
In the above we assume that the interface velocity
$\vel_i$ has been smoothly extended for all $x$.
Since the indicator function satisfies the so-called topological equation
(cf. \cite{Drew1999})
\begin{equation}
 \partial_t X_k + \vc{u}_i \cdot \nabla X_k = 0, \label{eqn:topological}
\end{equation} 
the first and the second term equations \eqref{pa1} and \eqref{pa2}
drop out, respectively, 
and we can write the system as
\begin{align}
 \partial_t \ensavg{X_k \rho} + \nabla \cdot \ensavg{X_k \rho \vel} &= \vc{\Gamma}_k \label{eqn:der_mass_source_def} \\
 \partial_t \ensavg{X_k \rho \vel} + \nabla \cdot \ensavg{X_k \rho \vel \otimes \vel} \quad \\- \nabla \cdot \ensavg{X_k \ts{T}} &=
 \ensavg{X_k \vc{f}} + \vc{M}_k \label{eqn:der_mom_source_def},
\end{align}
where 
\begin{align}
 \vc{\Gamma}_k &\equiv \ensavg{\rho (\vel - \vel_i) \cdot \nabla X_k}, \label{eqn:inter_mass_source_def} \\
 \vc{M}_k &\equiv \ensavg{\nabla X_k \cdot [\rho (\vel - \vel_i) \otimes \vel - \ts{T}]}, \label{eqn:inter_mom_source_def}
\end{align}
denotes the average interfacial mass source and the average interfacial momentum source for the $k$-th phase, respectively.

To obtain the averaged form of the jump conditions, 
we note first the Dirac delta property of the component indicator functions' derivative
\begin{equation}
 \ensavg{\nabla X_k f} = -\int_{\mathcal{S}_k} \vc{n}_k f_k d\vc{S},
\end{equation}
with $\mathcal{S}_k$ the interface of phase $k$. Using this and
\eqref{eqn:inter_mass_source_def}, \eqref{eqn:inter_mom_source_def} 
in the jump conditions for 
mass \eqref{eqn:jump_cond_def1} and momentum \eqref{eqn:jump_cond_def2},
these conditions become
\begin{align}
 \sum_k \vc{\Gamma}_k &= 0, \label{eqn:mass_jump_condition}\\
 \sum_k \vc{M}_k &= \vc{0}. \label{eqn:momentum_jump_condition}
\end{align}

We further introduce the following averaged quantities 
\begin{align*}
 \vf_k &\equiv \ensavg{X_k}, \\
 \intertext{for the volume fraction, and} 
 \pavg{\rho}_k &\equiv \frac{\ensavg{X_k \rho}}{\vf_k},  \\
 \favg{\vel}_k &\equiv \frac{\ensavg{X_k \rho \vel}}{\vf_k \pavg{\rho}_k},  \\
 \pavg{\ts{T}}_k &\equiv -\frac{\ensavg{X_k \ts{T}}}{\vf_k},  \\
 \ts{T}_k^{Re} &\equiv -\frac{\ensavg{X_k \rho \fder \vel_k \otimes \fder \vel_k}}{\vf_k}, \\
 \pavg{\vc{f}}_k &\equiv \frac{\ensavg{X_k \vc{f}}}{\vf_k},  \\
 \vc{S}_k^d &\equiv -\ensavg{\nabla X_k \cdot \ts{T}},  \\
 \pavg{\vel}_{ki} \vc{\Gamma}_k &\equiv \ensavg{\nabla X_k \cdot \rho (\vel - \vel_i) \otimes \vel}  
\end{align*} 
for the average density, velocity, stress, Reynolds stress, body forces, interfacial stress, interfacial velocity of the $k$th phase, respectively. 

Then, after we split the interfacial momentum source as
\begin{equation}\label{imss}
 \vc{M}_k = \vc{S}_k^d + \pavg{\vel}_{ki} \vc{\Gamma}_k 
\end{equation}
and the momentum flux into an average flux and
a Reynolds stress 
\begin{align}
 \ensavg{X_k \rho \vel \otimes \vel} =  \vf_k \pavg{\rho}_k \favg{\vel}_k \otimes \favg{\vel}_k - \vf_k \ts{T}_k^{Re}, 
\end{align}
and use the product rule \eqref{eqn:favre_product_rule} for the velocity, we obtain the following system of phase averaged mass and momentum equations 
\begin{align}
  \partial_t (\vf_k \pavg{\rho}_k) + \nabla \cdot (\vf_k \pavg{\rho}_k \favg{\vel}_k) &= \vc{\Gamma}_k, \\
 \partial_t (\vf_k \pavg{\rho}_k \favg{\vel}_k) + \nabla \cdot (\vf_k \pavg{\rho}_k \favg{\vel}_{k} \otimes \favg{\vel}_k) - \nabla \cdot (\vf_k \pavg{\ts{T}}_k) &= \\
  \nabla \cdot (\vf_k \ts{T}_k^{Re}) + \pavg{\vc{f}}_k + \vc{S}_k^d + \pavg{\vel}_{ki} \vc{\Gamma}_k.
\end{align}
The Reynolds stress $\ts{T}_k^{Re}$ consists of two parts - liquid turbulence and pseudo-turbulence. As we are interested in the laminar flow regime we neglect the liquid turbulence. Additionally, our derivations show $\fvel - \svel$ has a very small value in the considered flow cases. Since the pseudo-turbulence scales as $(\fvel - \svel)^2 \vf_s$, see e.g.~\cite{Cartellier2009, Fox2014}, it will also be neglected. Further, we assume no phase change occurs at the interface between particles and liquid, $\vc{\Gamma}_k = 0$.

We introduce the stress tensor as the sum of pressure and deviatoric stress in the form 
\begin{equation}
 \ts{T} = -p \ts{I} + \ts{\tau}, 
\end{equation}
so that for the averaged quantities $\pavg{\ts{T}}_k$ and
\begin{align}
\pavg{p}_k &\equiv \frac{\ensavg{X_k p}}{\vf_k},  \\
\pavg{\ts{\tau}}_k &\equiv -\frac{\ensavg{X_k \ts{\tau}}}{\vf_k},  
\end{align}
we have correspondingly 
\begin{equation}
 \pavg{\ts{T}}_k = -\pavg{p}_k \ts{I} + \pavg{\ts{\tau}}_k. 
\end{equation}

The interfacial pressure of phase $k$ and the
interfacial force density is defined as 
\begin{align}\label{pikintro}
\tilde{p}_{ik} &\equiv \frac{\ensavg{\nabla X_k p_{k}}}{\ensavg{\nabla X_k}} = \frac{\ensavg{\nabla X_k p_{k}}}{\nabla \vf_k},
\\
\vc{M}_k^d &\equiv \vc{S}_k^d - \ensavg{\nabla X_k p_{k}} = \ensavg{\nabla X_k \cdot ((p_k - \tilde{p}_{ik}) \ts{I} - \ts{\tau})},
\end{align}
respectively,
where the second equality in \eqref{pikintro} follows from an application of 
Gauss' rule \eqref{eqn:GaussRule}. 
We have (from \eqref{imss})
\begin{equation}
 \vc{M}_k = \vc{M}_k^d + \tilde{p}_{ik} \nabla \vf_k,
\end{equation}
so that we obtain for the mass and momentum balance equations
\begin{align}
  \partial_t (\vf_k \pavg{\rho}_k) + \nabla \cdot (\vf_k \pavg{\rho}_k \favg{\vel}_k) &= 0, \\
\partial_t (\vf_k \pavg{\rho}_k \favg{\vel}_k) + \nabla \cdot (\vf_k \pavg{\rho}_k \favg{\vel}_k \otimes \favg{\vel}_k) \quad \\
 - \nabla \cdot (\vf_k \pavg{\ts{\tau}}_k) + \nabla (\vf_k \pavg{p}_k) &= \vc{M}_k^d + \tilde{p}_{ik} \nabla \vf_k,
\end{align}
where we have also assumed that no external body forces are applied, i.e. $\pavg{\vc{f}}=0$.

We neglect surface tension forces between the solid and the liquid phase, so the interfacial pressure difference becomes \cite{Drew1983}
\begin{align}
 \sum_k \tilde{p}_{ik} \nabla \vf_k = 0,
 \label{eqn:der_pik_surface}
\end{align}
and we obtain together with the interfacial momentum jump condition 
\eqref{eqn:momentum_jump_condition} the relation
\begin{equation}
 M_s^d = -M_f^d.
\end{equation}
Since we only have two phases, we know
$\vf_s + \vf_f = 1,$
which directly leads to $\nabla \vf_s = -\nabla \vf_f$. Thus, equation \eqref{eqn:der_pik_surface} yields
\[
 \tilde{p}_{is} = \tilde{p}_{if}.
\]
For the case of identical liquid interfacial and bulk pressure 
\[ 
\tilde{p}_{if} = \pavg{p}_f,
\]
and
constant densities $\pavg{\rho}_k$  within each phase,
the balance equations then reduce to the system \eqref{eqn:govern_dimens_system}.

\section{Boundary layer analysis for the drift-flux model}\label{boundary_layer_appendix}

In this appendix, we complete the perturbation analysis used for the derivation
of the drift-flux model in section~\ref{subsec:ad} by considering the inner
layer near the wall. The purpose of this is to show that we recover the no-flux
condition $w_2=0$ used to complete the drift-flux model \eqref{driftfluxout},
but we note that for a correct description of the density profile in the inner
layer, which has a width $\varepsilon^{1/2}=$Da$^{-1/2}\sim K_p^{1/2}\sim a$
i.e.\ of the size of the particles, we would have to include the possibility
of a depletion layer, which, however, should not affect the no-flux condition 
on \eqref{driftfluxout}.

For the boundary layer analysis at the wall we introduce variable 
\be
z=\frac{\frac12-y}{\eps^{1/2}}, \qquad \Phi(t,z)=\phi(t,y).
\ee
Then we obtain 
\begin{subequations}
\begin{align}
\eps^{1/2}\pt\Phi + \pz(\Phi\,(1- \Phi)\, w_2) &= 0 \\
-\pz\left[(1-\Phi)\,\pz v_1 + \eps(1-\Phi)\,\pz(\Phi w_1)\right] +\eps(1-\Phi)\px p_f &= -\eps\frac{\Phi^2}{1-\Phi} w_1\\
\eps^{1/2}\pz\left[2(1-\Phi)\,\pz(\Phi w_2)\right] + (1-\Phi)\pz p_f &= \eps^{1/2}\frac{\Phi^2}{1-\Phi} w_2 \\
-\pz\left[\Phi\eta_s\pz v_1 - \eps\Phi\eta_s\pz((1-\Phi) w_1)\right] + \eps\Phi\px p_f &= \eps\frac{\Phi^2}{1-\Phi} w_1 \\
\eps^{1/2}\pz\left[2\Phi\,\pz((1-\Phi) w_2)\right] - \Phi \pz p_f - \pz p_c &= \eps^{1/2}\frac{\Phi^2}{1-\Phi} w_2 
\end{align}
and 
\be
p_c =\eta_n\left[\frac{1}{\eps}(\pz v_1 - \eps\pz((1-\Phi) w_1))^2 + 2[\pz((1-\Phi) w_2)]^2\right]^{1/2}
\ee
and no-slip conditions at $z=0$
\be
v_1=0,\quad w_1=0,\quad w_2=0.
\ee
\end{subequations}
The leading order system is
\begin{subequations}
\begin{align}
 \pz(\Phi\,(1- \Phi)\, w_2) &= 0 \\
-\pz\left[(1-\Phi)\,\pz v_1 \right]  &= 0\\
(1-\Phi)\pz p_f &= 0 \\
-\pz\left[\Phi\eta_s\pz v_1 \right]  &= 0 \\
-\pz p_c &= 0
\end{align}
and 
\be
p_c=\eta_n\left[ (\pz v_1)^2 \right]^{1/2}
\ee
and no-slip conditions at $z=0$
\be
v_1=0,\quad w_1=0,\quad w_2=0.
\ee
\end{subequations}

We see immediately that $w_2=0$, which provides, via matching, the boundary condition for the drift-flux model at $y=1/2$ as claimed in the text.

\end{appendix}


\bibliography{driftfluxjam}
\bibliographystyle{abbrv}


\label{lastpage}

\end{document}